\documentclass[12pt]{iopart}

\usepackage{iopams}
\usepackage{graphicx}
\usepackage{bm}
\usepackage{subfigure}
\usepackage{xcolor}
\begin{document}

\title[Resolving the fine-scale structure in turbulent Rayleigh-B\'enard convection]{Resolving the fine-scale structure in turbulent Rayleigh-B\'enard convection}

\author{Janet D. Scheel}
\address{Department of Physics, Occidental College, Los Angeles, CA 90041, USA}
\ead{{jscheel@oxy.edu}}

\author{Mohammad S. Emran}
\address{Institut f\"ur Thermo- und Fluiddynamik, Postfach 100565, Technische Universit\"at Ilmenau,
         D-98684 Ilmenau, Germany}
\ead{{mohammad-shah.emran@tu-ilmenau.de}}

\author{J\"org Schumacher}
\address{Institut f\"ur Thermo- und Fluiddynamik, Postfach 100565, Technische Universit\"at Ilmenau,
         D-98684 Ilmenau, Germany}
\ead{{joerg.schumacher@tu-ilmenau.de}}

\begin{abstract}
We present high-resolution direct numerical simulation studies of turbulent Rayleigh-B\'{e}nard convection in a 
closed cylindrical cell  with an aspect ratio of one. The focus of our analysis is on the finest scales of  convective turbulence, in 
particular the statistics of the kinetic energy and thermal dissipation rates in the bulk and the whole cell. The fluctuations of the 
energy dissipation field can directly be translated into a fluctuating local dissipation scale which is 
found to develop ever finer fluctuations with increasing Rayleigh number. The range of these scales as well as the
probability of high-amplitude dissipation events decreases with increasing Prandtl number. In addition, we
examine the joint statistics of the two dissipation fields and the consequences of high-amplitude events. 
We also have investigated the convergence properties of our spectral element method and have found that both dissipation 
fields are very sensitive to insufficient resolution. We demonstrate that global transport properties, such as the Nusselt number,  
and the energy balances are partly insensitive to insufficient resolution and yield correct results even when the dissipation 
fields are under-resolved. Our present numerical framework is also compared with high-resolution simulations which use 
a finite difference method. For most of the compared quantities the agreement is found to be satisfactory.
\end{abstract}

%\submitto{\NJP}
%\tableofcontents
\maketitle

\section{Introduction}
Turbulent fluid motion in nature and technology is frequently driven by sustained temperature
differences \cite{Chilla2012}. Applications range from cooling devices of chips to convection in the Earth and
the Sun. Turbulent Rayleigh-B\'enard convection (RBC) is the paradigm for all these convective
phenomena because it can be studied in a controlled manner, but it still has enough
complexity to contain the key features of convective turbulence in the examples just
mentioned. RBC in cylindrical cells has been studied intensely over the last few years in
several laboratory experiments, mostly in slender cells of aspect ratio smaller than or equal to
unity in order to reach the largest possible Rayleigh numbers or to resolve the detailed mechanisms 
of turbulent heat transport close to the walls \cite{Ahlers2009,Chilla2012}.  Direct numerical simulations have also grown
such that the detailed dynamical and statistical aspects of the involved turbulent fields and their characteristic 
structures can now be unraveled in detail.

The key question in RBC is the mechanism of turbulent transport of heat and momentum. Since the 
fluid motion is driven by a constant temperature difference between the top and bottom plates, thin 
boundary layers of temperature and velocity will form on these walls as well as on the side walls of the cell. 
A deeper understanding of the global transport mechanisms is possible only if we understand
the dynamical coupling between the boundary layers and the rest of the flow in the bulk of the
cell. While the boundary layers are strongly dominated by the presence of mean gradients of the temperature and
velocity fields, the bulk of the convection layer is well mixed by the turbulence such that mean 
gradients of the involved turbulent fields remain subdominant compared to the local fluctuations. The flow 
at hand is thus strongly inhomogeneous, at least in the vertical direction, so it can be expected that the 
smallest dynamically relevant scales will differ when moving from the isothermal walls to the bulk. 

Central to our understanding of the statistics of the turbulent transport is the role of the gradient fields of 
velocity and temperature which fluctuate extremely strongly at the small scales of the flow. This is a 
unique property of all turbulent flows. Dissipation rate fields -- which measure the magnitude of these gradient fluctuations 
and are still inaccessible in experiments with respect to their three-dimensional structure \cite{Wallace2009}  -- are 
thus at the core of a deeper understanding of turbulence as a whole.

In the present work we want to make a further step forward with DNS of RBC by resolving fine scales never
accessed before, both in the bulk and boundary layers, in order to study the statistics of the gradient fields, their joint extreme events,  the statistical effect of rare high-amplitude events as well as  Rayleigh and  
Prandtl number variation. A spectral element method is used to conduct the 
numerical studies \cite{nek5000,Fischer1997}. It combines the flexibility in terms
of mesh geometry that is inherent to every finite element method with the exponentially fast convergence of a spectral 
method. We will show that several tests which have been applied in DNS in the past are insensitive with respect to 
insufficient resolution. These tests, which are based on global averages of mean dissipation rates and the global 
mean heat fluxes, give correct results although the fine-scale structure of the turbulence is still under-resolved particularly 
in the bulk of the convection cell.  In order to address these questions in detail we will present a comprehensive statistical 
analysis of the temperature and velocity gradient fields, in particular the related dissipation rates and dissipation scales.  

It is crucial to resolve all the dynamically important scales to represent the flow faithfully when carrying 
out Direct Numerical Simulations (DNS) which involve no subgrid-scale parametrization. 
Several attempts have been made in order to  derive resolution criteria starting with the pioneering work 
by Gr\"otzbach \cite{Groetzbach1983}, subsequent refinements of this criterion \cite{Stevens2010,Bailon2010,Shishkina2010} 
and works with a focus to the fine resolution of the boundary layer dynamics \cite{Wagner2012,Shi2012,Scheel2012}. 
 Only recently the focus of DNS studies was shifted towards the bulk in a 
cubic convection cell \cite{Kaczorowski2013} with a discussion of the scaling properties and statistics of the 
dissipation fields.

It is well-known that the gradients of the turbulent fields are most sensitive to insufficient resolution.
Superfine resolution simulations in isothermal box turbulence \cite{Schumacher2007,Schumacher2007a,Schumacher2010} and
in turbulent shear flows \cite{Hamlington2012} have led to some enlightening results on the distribution of the finest scales
in such flows and their relation to the small-scale intermittency. This intermittency is known to be coupled tightly to two 
highly fluctuating dissipation rates, one of the kinetic energy and the other of the thermal variance. The thermal dissipation rate 
is defined as
%-------------------------------------------------------------------------------
\begin{equation}
\epsilon_T({\bf x}, t)=\kappa (\bm{\nabla} T)^2\,,
\label{thermal}
\end{equation}
%-------------------------------------------------------------------------------
where $T(\bm x,t)$ is the temperature field and $\kappa$ the thermal diffusivity. The kinetic energy dissipation rate is defined as 
%-------------------------------------------------------------------------------
\begin{equation}
\epsilon({\bf x}, t)=\frac{\nu}{2} \left(\bm\nabla \bm u+\bm\nabla \bm u^T\right)^2\,,
\label{energy}
\end{equation}
%-------------------------------------------------------------------------------
with the turbulent velocity field $\bm u({\bf x},t)$ and the kinematic viscosity $\nu$. The mean kinetic energy dissipation 
rate $\langle\epsilon\rangle$ is related to the mean Kolmogorov scale $\langle \eta_K\rangle$ which is the smallest mean 
scale when $\nu\le\kappa$. The symbol $\langle\cdot\rangle$ denotes an ensemble average which is calculated in numerical
simulations as a volume-time average. In case of $\nu>\kappa$, the smallest mean scale is determined by the (active) scalar
field known as the Batchelor scale \cite{Batchelor1959}, $\langle \eta_B\rangle$. Both scales are defined as
%-------------------------------------------------------------------------------
\begin{equation}
\langle\eta_K\rangle=\frac{\nu^{3/4}}{\langle\epsilon\rangle^{1/4}} \;\;\;\;\mbox{and}\;\;\;\;\langle\eta_B\rangle=
\frac{\langle \eta_K\rangle}{\sqrt{Pr}}\,.
\label{diss1}
\end{equation}
%-------------------------------------------------------------------------------
Here, $Pr$ is the Prandtl number and given by
%-------------------------------------------------------------------------------
\begin{equation}
Pr=\frac{\nu}{\kappa}\,.
\end{equation}
%-------------------------------------------------------------------------------
Both dissipation fields can be expected to fluctuate strongly exceeding their means by orders of magnitude 
\cite{Sreenivasan1997,Schumacher2007}. Therefore it was suggested to generalize the classical dissipation 
and diffusion scales to local dissipation and diffusion scales \cite{Sreenivasan2004,Schumacher2005} 
which are given by (see also a discussion in Hamlington et al. \cite{Hamlington2012})
%-------------------------------------------------------------------------------
\begin{equation}
\eta_K({\bf x},t)=\frac{\nu^{3/4}}{\epsilon({\bf x},t)^{1/4}} \;\;\;\;\mbox{and}\;\;\;\;\eta_B({\bf x},t)=\frac{\eta_K({\bf x},t)}{\sqrt{Pr}}\,.
\label{diss1a}
\end{equation}
%-------------------------------------------------------------------------------
Both scales will pick up the highly intermittent fluctuations of the dissipation rates and can thus become smaller, but also larger
than the mean scales which were defined in (\ref{diss1}). Local dissipation scales have been studied in convection experiments
by Zhou and Xia \cite{Zhou2010}. One main finding was that the distribution of the scales can be described by the same tools
as in isothermal box turbulence. In the present work we will also access these scales and compare their distribution in different 
parts of the convection cell.

The manuscript is organized as follows. In Sec. 2, we will discuss in brief the equations of motion and some central 
relations which will become necessary for our data analysis. Furthermore we briefly review the existing resolution criteria. First it is shown that global balance equation checks are insensitive to insufficient resolution. 
We also compare the results with a 
second-order finite difference method \cite{Verzicco1996,Verzicco2003} which has been one workhorse in RBC over the past decade. 
In Sec. 3 we report our results. We study the statistics of the dissipation rate fields and calculate the local dissipation scales. 
Then we present a comparison of dissipation rate fields and scales as a function of Rayleigh number and Prandtl number using our 
very highest resolution.  We summarize our findings and give a brief outlook at the end.

\section{Equations of motion and numerical method}
\subsection{Boussinesq equations and further non-dimensional relations}
We solve the three-dimensional Boussinesq equations numerically. The height of the cell $H$, the free-fall velocity 
$U_f=\sqrt{g \alpha \Delta T H}$ and the imposed temperature difference $\Delta T$ are used to rescale the equations 
of motion. The three control parameters of Rayleigh-B\'enard convection are the Rayleigh number $Ra$, 
the Prandtl number $Pr$ and the aspect ratio $\Gamma=D/H$ with the diameter $D$. This results in the following dimensionless form 
of the equations of motion
%-------------------------------------------------------------------------------
\begin{eqnarray}
\label{ceq}
& &\tilde {\bm \nabla}\cdot\tilde {\bf u}=0\,,\\
\label{nseq}
& &\frac{\partial\tilde{\bf u}}{\partial \tilde t}+(\tilde{\bf u}\cdot\tilde{\bm\nabla})\tilde{\bf u}
=-\tilde{\bm \nabla} \tilde p+\sqrt{\frac{Pr}{Ra}} \tilde{\bm \nabla}^2\tilde{\bf u}+ \tilde T {\bf e}_z\,,\\
& & \frac{\partial \tilde T}{\partial \tilde t}+(\tilde {\bf u}\cdot\tilde {\bm \nabla}) \tilde T
=\frac{1}{\sqrt{Ra Pr}} \tilde{\bm \nabla}^2 \tilde T\,,
\label{pseq}
\end{eqnarray}
%-------------------------------------------------------------------------------
where
%-------------------------------------------------------------------------------
\begin{equation}
Ra=\frac{g\alpha\Delta T H^3}{\nu\kappa}\,.
\end{equation}
%-------------------------------------------------------------------------------
The variable $g$ stands for the  acceleration due to gravity and $\alpha$ is the thermal expansion coefficient.
Throughout the study we set $\Gamma=1$. Times are measured in free-fall time units, $T_f=\sqrt{H/(g\alpha\Delta T)}$.
At all walls of the simulation volume $V$ no-slip boundary conditions for the 
fluid are applied, $\tilde{\bf u}=0$. The side walls are adiabatic, i.e., the normal derivative of the temperature field vanishes,
$\partial \tilde T/\partial {\bf n}=0$. The top and bottom plates are held at  fixed temperatures $\tilde T=0$ and 1, respectively. In response to
the input parameters $Ra$, $Pr$ and $\Gamma$, a turbulent heat flux from the bottom to the top plate is established.  
It is determined by the Nusselt number which is defined as  
%-------------------------------------------------------------------------------
\begin{equation}
Nu(\tilde z)=\sqrt{Ra Pr}\, \langle\tilde u_z \tilde T\rangle_{A,t}-\frac{\partial\langle \tilde T\rangle_{A,t}}{\partial \tilde z}\,.
\end{equation}
%-------------------------------------------------------------------------------
Based on the volume average, we find $Nu_V=1+\sqrt{Ra Pr}\langle \tilde u_z \tilde T\rangle_{V,t}$ which has to equal $Nu(\tilde z)$ 
for all $\tilde z\in [0,1]$.
The non-dimensional expressions for the two dissipation rate fields, $\epsilon(\bm x,t)$ and $\epsilon_T(\bm x,t)$ are given by the following expressions:
%-------------------------------------------------------------------------------
\begin{equation}
\epsilon_T({\bf x}, t)=\kappa \frac{(\Delta T)^2}{H^2}  (\tilde{\bm \nabla} \tilde{T})^2=
\frac{(\Delta T)^2 U_f}{H} \frac{1}{\sqrt{Ra Pr}}(\tilde{\bm \nabla} \tilde{T})^2\,,
\label{thermal_0}
\end{equation}
%-------------------------------------------------------------------------------
and thus
%-------------------------------------------------------------------------------
\begin{equation}
\tilde{\epsilon}_T(\tilde{\bf x},\tilde t) \equiv {\epsilon_T({\bf x}, t) H\over U_f(\Delta T)^2} =
\frac{1}{\sqrt{Ra Pr}}(\tilde{\bm \nabla} \tilde{T})^2\,.
\label{thermal1}
\end{equation}
%-------------------------------------------------------------------------------
The kinetic energy dissipation rate is defined as 
%-------------------------------------------------------------------------------
\begin{equation}
\epsilon({\bf x}, t)=\frac{\nu}{2}\, \frac{U_f^2}{H^2}  \left(\tilde{\bm\nabla} \tilde{\bm u}+\tilde{\bm\nabla} \tilde{\bm u}^T\right)^2=
                                 \frac{U_f^3}{2H} \sqrt{\frac{Pr}{Ra}} \left(\tilde{\bm\nabla} \tilde{\bm u}+\tilde{\bm\nabla} \tilde{\bm u}^T\right)^2\,,
\label{kinetic}
\end{equation}
%-------------------------------------------------------------------------------
and thus
%-------------------------------------------------------------------------------
\begin{equation}
\tilde\epsilon(\tilde{\bf x}, \tilde t) \equiv {\epsilon({\bf x}, t) H\over U_f^3} =\frac{1}{2} \sqrt{\frac{Pr}{Ra}}\left(\tilde{\bm\nabla} \tilde{\bm u}+\tilde{\bm\nabla} \tilde{\bm u}^T\right)^2\,.
\label{kinetic1}
\end{equation}
%-------------------------------------------------------------------------------
Using equation (\ref{diss1a}) gives
%-------------------------------------------------------------------------------
\begin{equation}
\tilde\eta_K(\tilde{\bf x}, \tilde t) 
\equiv {\eta_K(\tilde {\bf x}, \tilde t)\over H} = 
\left [\frac{\nu^{3/4}}{(U_f H)^{3/4}}\right ]\,\tilde\epsilon(\tilde{\bf x},\tilde t)^{-1/4}=\frac{Pr^{3/8}}{Ra^{3/8}}\,\tilde\epsilon(\tilde{\bf x},\tilde t)^{-1/4}\,,
\label{energy2}
\end{equation}
%-------------------------------------------------------------------------------
for the cases of $Pr\le 1$. To simplify (\ref{energy2}) we used the definition of the Grashof number $Gr=(U_f H)^2/\nu^2=Ra/Pr$. The Batchelor scale follows as
%-------------------------------------------------------------------------------
\begin{equation}
\tilde\eta_B(\tilde{\bf x}, \tilde t) \equiv {\eta_B(\tilde{\bf x}, \tilde t)\over H} = \frac{1}{Pr^{1/8}Ra^{3/8}}\,\tilde\epsilon(\tilde{\bf x},\tilde t)^{-1/4}\,.
\label{energy2a}
\end{equation}
%-------------------------------------------------------------------------------
for $Pr>1$. For completeness, we also list two exact relations that can be derived from the balances of the turbulent kinetic energy and the scalar 
variance. They are given by \cite{Shraiman1990}
%-------------------------------------------------------------------------------
\begin{equation}
\langle \tilde\epsilon_T\rangle_{V,t}=\frac{Nu}{\sqrt{Ra Pr}}\;\;\;\;\mbox{and}\;\;\;\;\langle \tilde\epsilon\rangle_{V,t}=\frac{Nu-1}{\sqrt{Ra Pr}}.
\label{rel1}
\end{equation}
%-------------------------------------------------------------------------------
If we make use of (\ref{rel1}), Eqns. (\ref{energy2}), and (\ref{energy2a}) translate to
%-------------------------------------------------------------------------------
\begin{equation}
\langle\tilde{\eta}_K\rangle=\left(\frac{Pr^2}{(Nu-1) Ra}\right)^{1\over 4} \;\;\;\;\mbox{and}\;\;\;\;\langle\tilde{\eta}_B\rangle=
\left(\frac{1}{(Nu-1) Ra}\right)^{1\over 4}\,.
\label{diss1b}
\end{equation}
%-------------------------------------------------------------------------------
Similar to the studies by Stevens et al. \cite{Stevens2010}, we will use eqns. (\ref{rel1}) to test different grid resolutions at a given set 
of parameters and define relative errors that measure the difference between the left and right hand sides of Eqns. (\ref{rel1})
\footnote{Note that the relative errors are fairly sensitive to the averaging time because of the large fluctuations in Nusselt number 
that can occur for these turbulent systems.}
%-------------------------------------------------------------------------------
\begin{equation}
\Lambda_T=\frac{\sqrt{Ra Pr}\langle \tilde\epsilon_T\rangle_{V,t}-Nu}{Nu}\;\;\;\;\mbox{and}\;\;\;\;\Lambda_v=\frac{\sqrt{Ra Pr}\langle \tilde\epsilon\rangle_{V,t}-(Nu-1)}{Nu-1}\,.
\label{rel1a}
\end{equation}
%-------------------------------------------------------------------------------
In the following, we will continue with the dimensionless quantities and omit the tildes for convenience.

\subsection{Numerical methods}
For the DNS studies in the present work two different numerical methods are used and compared, a second-order finite difference scheme and
a spectral element method.
  
\subsubsection{Finite difference method}
The Boussinesq equations (\ref{ceq})--(\ref{pseq}) are discretized on a staggered grid with a second-order finite 
difference scheme (FDM) which was developed by Verzicco and Orlandi \cite{Verzicco1996,Verzicco2003}. The pressure 
field $p$ is determined by a two-dimensional Poisson solver after applying a one-dimensional Fast Fourier Transform (FFT) 
in the azimuthal direction. The time advancement is done by a third-order Runge-Kutta scheme. The grid spacings are 
non-equidistant in the radial and  vertical directions. In the vertical direction, the grid spacing is close  to Tschebycheff 
collocation points. The simulation code is parallelized with MPI in combination with OpenMP.

\subsubsection{Spectral element method}
The equations are numerically solved by a spectral element method (SEM) \cite{nek5000} which has been adapted to our 
problem. The code employs second order time-stepping, using the backward difference formula BDF2 which results 
at time step $n$ and for a step width $\delta t$ in the following set of discrete equations (see also Eqns. (\ref{ceq})--(\ref{pseq}))
%-------------------------------------------------------------------------------
\begin{eqnarray}
& &\hat D\bm u^n=0\,, \label{rel0ab}\\
& &\sqrt{\frac{Pr}{Ra}}\hat{A} {\bm u}^n +\frac{3}{2\delta t}\hat{B} {\bm u}^n + \bm\nabla p^n= \bm f_u^n\,,\\
& &\frac{1}{\sqrt{Ra Pr}}\hat{A} T^n +\frac{3}{2\delta t}\hat{B} T^n =  f_T^n\,,
\label{rel1ab}
\end{eqnarray}
%-------------------------------------------------------------------------------
with the corresponding boundary conditions. Here, $\hat{D}$ is the divergence operator and $\hat A$ the stiffness matrix which contains 
the Laplace terms. The quantity $\hat B$ is the mass matrix which will contain the Gauss-Lobatto-Legendre weights and the determinants 
of the Jacobian caused by the mapping to the deformed elements as diagonal entries.
In order to arrive at (\ref{rel0ab})--(\ref{rel1ab}) the Boussinesq equations are transformed into a weak formulation similar to other 
Galerkin methods. They are then discretized with the particular choice of spectral basis functions \cite{Deville2002} which will be given further below. 
These basis functions allow for an exact evaluation of the integrals in the scalar products on the basis of the Gauss integration theorem.  
All flow fields are given in the Sobolev space $H^1(V)$ in which the functions and their derivatives are square integrable. For this space it holds
that $C^1(V)\subset H^1(V)\subset C^0(V)$ \cite{Deville2002}. The right hand sides, 
$\bm f_u^n$ and $f_T^n$ of Eqns. (\ref{rel1ab}), incorporate remaining  terms from the BDF2 time derivative,  the nonlinear convection which 
is  obtained by second order extrapolation from steps $n-1$ and $n-2$, and the buoyancy. 

The resulting linear, symmetric Stokes problem is solved implicitly. This system is split, decoupling the viscous and pressure
steps into independent symmetric positive definite subproblems which are solved either by Jacobi (viscous) or 
multilevel Schwartz (pressure) preconditioned conjugate gradient iteration. Fast parallel solvers based on direct projection \cite{Tufo2001} or 
more scalable algebraic multigrid \cite{Fischer2008} are used for the coarse-grid solve that is part of the pressure preconditioner. For stabilization of the SEM, we perform de-aliasing by the use of over-integration of the convective term by a factor of either $N+5$ or $3(N+1)/2$, where $N$ is the polynomial order. We also filter out 5\% of the energy in the $N$th mode for additional stabilization (see \cite{Fischer2001} for further information).

The basis functions $\psi_k(x)$ are Lagrangian interpolation polynomials of order $N$ and composed of  Legendre polynomials $P_k$ 
for the present study.  They are given by
%----------------------------------------------------------------------
\begin{equation}
\psi_k(x)= -\frac{1}{N(N+1)}\,\frac{(1-x^2) P_N^{\prime}(x)}{(x-\xi_k)P_{N}(\xi_k)}\,,
\label{basefunctions}
\end{equation}
%----------------------------------------------------------------------
with the Gauss-Lobatto-Legendre points $\xi_k$. In contrast to a classical
spectral or pseudospectral method the evaluation of spatial derivatives translates into matrix multiplications which have to be 
highly optimized (see the appendix for further details).  The expansion in the three-dimensional case with a 
reference element $\Omega=[-1, 1]^3$ is based on the tensor product formulation of the basis functions 
%----------------------------------------------------------------------
\begin{equation}
{\bf u}_e(x,y,z)=\sum_{i=0}^N \sum_{j=0}^N\sum_{k=0}^N{\bf u}(\xi_i,\xi_j,\xi_k) \;\psi_i(x)\!\otimes\!\psi_j(y)\!\otimes\!\psi_k(z)\,.
\label{expans2}
\end{equation}
%----------------------------------------------------------------------
In the simulation the elements that sum up to the volume $V$ are deformed. Hence an additional mapping (Jacobian) from the reference element 
to all elements needs to be incorporated. Clearly,  the mapping of the coordinates and the matching of the velocity and temperature
fields between elements enhances the numerical effort in comparison to the second order FDM. We estimated that production runs on the same number 
of cores for the same system size would be approximately 10 times slower. In turn, gradient fields are calculated on each element separately with
an exponentially fast convergence.
 
%---------------------------------------------------------------------
\begin{figure}
\centering
\includegraphics[width=0.46\textwidth]{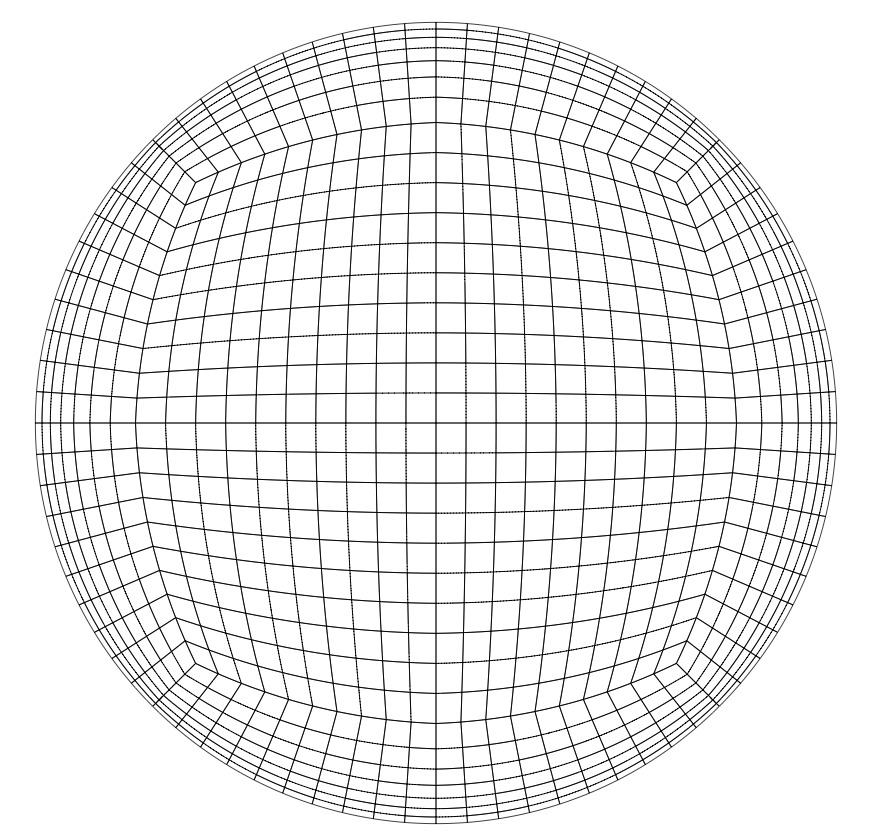}
\includegraphics[width=0.46\textwidth]{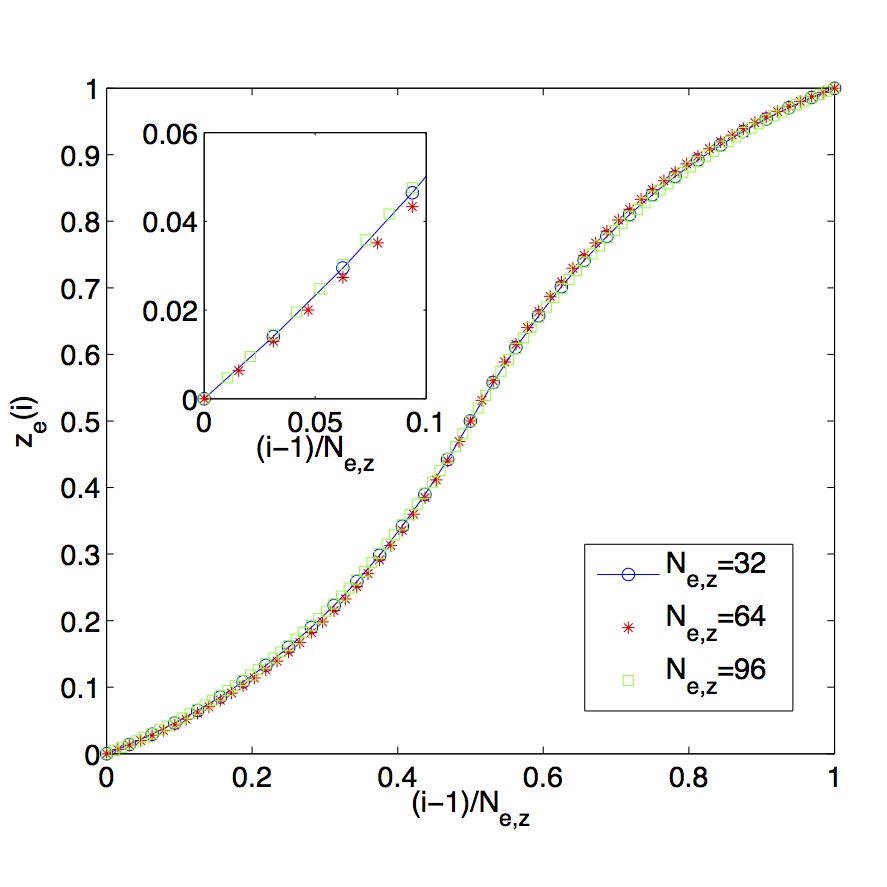}
\caption{Left: Display of the horizontal primary node structure as used for runs SEM1 to SEM4 displayed in Table \ref{Tab1}. 
Right: Display of the vertical primary node mesh for runs SEM6 to SEM9 (see Table \ref{Tab2}). The stretching factors $r$ are  $r=0.91$ 
for $N_{e,z}=32$, $r=0.95$ for $N_{e,z}=64$, and $r=0.97$ for $N_{e,z}=96$.}
\label{grid_0}
\end{figure}
%---------------------------------------------------------------------

\subsection{Existing resolution criteria for direct numerical simulations}
The first estimation of spatial resolution requirements for direct numerical simulations of Rayleigh-B\'enard convection were made by Gr\"otzbach
\cite{Groetzbach1983}. His criteria for confined convection cells consisted of (I) resolving the steep gradients in the velocity and temperature near 
the walls with a sufficient vertical grid width distribution and (II) resolving the smallest relevant turbulence elements with a sufficiently small mean 
grid width. Based on tests of Nusselt number with a spectral code, Gr\"otzbach's first criterion requires at least 3 nodes within the thermal boundary 
layer thickness for Prandtl numbers on the order of one or larger. For much smaller Prandtl numbers,  more nodes may be necessary as the viscous 
boundary layer becomes much thinner than the thermal boundary layer. The second criterion translates to a relation between the mean grid width 
$\bar{\Delta}$ and the mean dissipation or diffusion scale. For $Pr \leq 1$ this relation is $\bar{\Delta} \leq \pi\langle \eta_K\rangle$ and for $Pr \geq 1$ it is   
$\bar{\Delta} \leq \pi\langle \eta_B\rangle$. Gr\"otzbach then assumes that $\langle \eta_K\rangle$  and  $\langle \eta_B\rangle$ can be approximated 
by the mean kinetic energy dissipation rate $\langle\epsilon\rangle$ as in Eq. (\ref{diss1}). By using an argument similar to Eqns. (\ref{rel1}) this 
leads to the following global criteria on the grid widths\footnote{The factor of $\pi$ can be rationalized to our view by the resolution criteria as 
formulated for pseudospectral box turbulence simulations (see e.g. \cite{Schumacher2005}). There, $k_{max}\langle\eta_K\rangle\ge 1$ should be
satisfied with the maximum resolved wave number (after de-aliasing) $k_{max}=\sqrt{2}N_x/3$ and $N_x=N_y=N_z$ equals the number of grid points 
in each direction. The standard box length is then the periodicity length of the Fourier modes $L_x=2\pi$ and thus $N_x=2\pi/\bar{\Delta}$.}:
%-------------------------------------------------------------------------------
\begin{eqnarray}
\bar{\Delta} \leq \pi \left({Pr^2\over Ra (Nu-1)}\right)^{1\over 4} &\ {\rm for}\ & Pr \leq 1, \\
\label{gbsle1}
\bar{\Delta} \leq \pi \left({1\over Ra (Nu-1)}\right)^{1\over 4} &\ {\rm for}\ & Pr \geq 1. 
\label{gbsle2}
\end{eqnarray}
%-------------------------------------------------------------------------------

The criteria of Gr\"otzbach were revised by Stevens et al. \cite{Stevens2010}  based on DNS results using the second order finite difference 
method also used for comparison in this paper \cite{Verzicco1996,Verzicco2003}. They systematically found the Nusselt number to be overestimated 
in poorly resolved simulations, especially when the plume dynamics were not properly resolved. They suggested changing the {\it mean} grid width 
criteria to one that instead holds for the {\it largest} grid width in any spatial dimension, since the Kolmogorov length needs to always be resolved in 
order to properly characterize the flow. A similar perspective was developed in Bailon-Cuba et al. (2010). Although Stevens et al. (2010) did not 
determine any exact resolution criteria, they did compute the volume averaged dissipation rates 
$\langle \epsilon\rangle$ and $\langle \epsilon_T\rangle$ and compared these values to the globally computed Nusselt number as we have done in 
equation (\ref{rel1}). They found that for high enough Rayleigh number ($\geq 10^9$), even though the Gr\"otzbach criteria was technically followed, and 
equation (\ref{rel1}) was well-satisfied for the viscous dissipation rate, equation (\ref{rel1}) was not as well-satisfied for the thermal dissipation rate.

A further revision was conducted by Shishkina et al. \cite{Shishkina2010},  who used the theoretical Prandtl-Blasius (PB) theory to derive 
a lower bound on the number of nodes required to be placed in both the thermal and the viscous boundary layers such that the estimated Kolmogorov 
lengths in the boundary layers are adequately resolved. For higher Rayleigh number, this minimum bound is much larger than that suggested by 
Gr\"otzbach. For example, for our parameter range ($\Pr = 0.7$), Shishkina et al. suggest a minimum of 5 nodes for $Ra = 2\times 10^7$  but 
increasing to 9 nodes for $Ra = 2\times 10^9$. 

We will also discuss our own results in light of these criteria, including Gr\"otzbach and the revisions by Stevens and Shishkina. However, we will 
take this analysis one step further by investigating the implications of resolving not only  the global but also the local dissipation scales.
%---------------------------------------------------------------------
\begin{figure}
\centering
\includegraphics[width=1.0\textwidth]{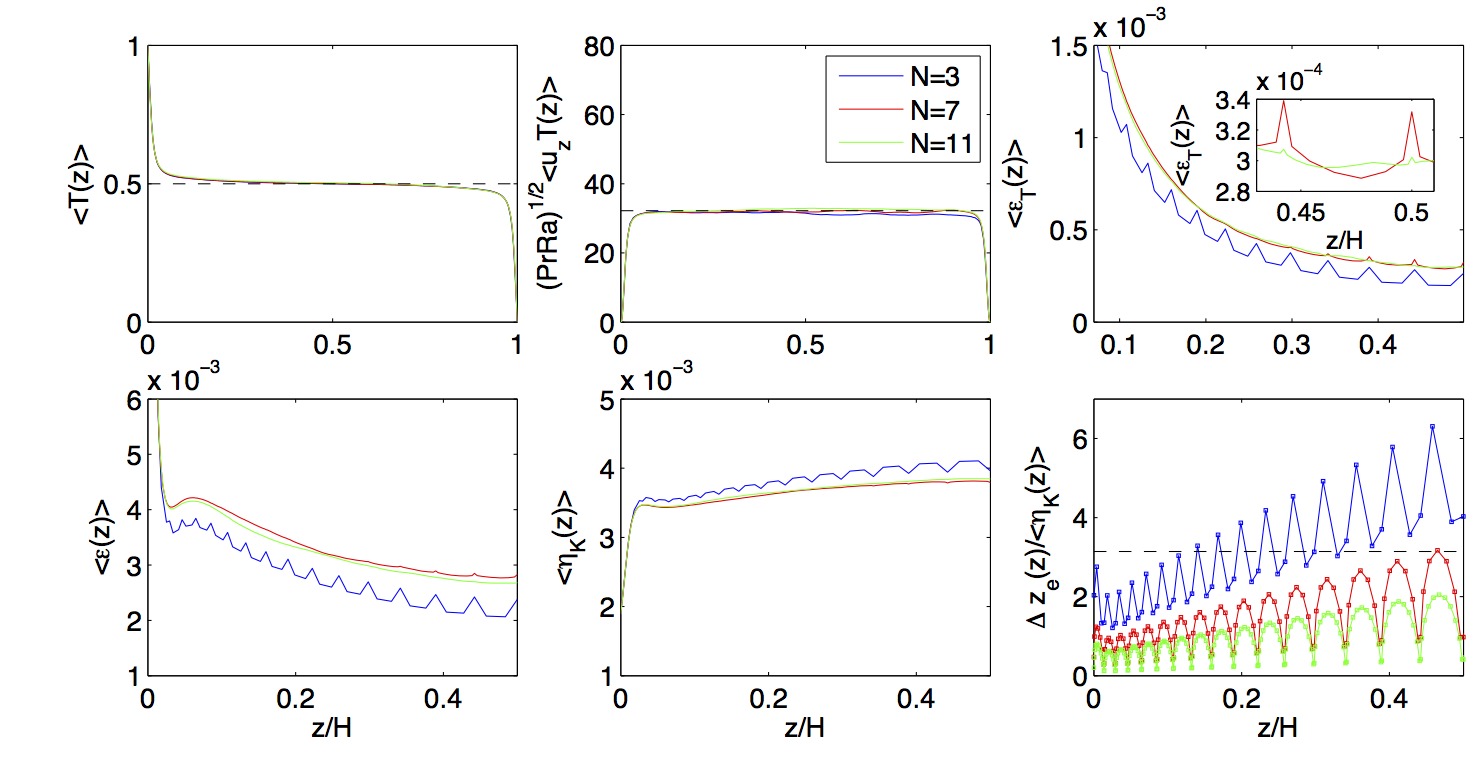}
\caption{Resolution tests for $Ra=10^8$ and $N=3, 7$ and 11 using the same primary mesh (runs SEM1, SEM3 and SEM4 in Table \ref{Tab1}).
We compare the mean temperature profile $\langle T(z)\rangle_{A,t}$, the mean convective flux profile $\sqrt{Ra Pr}\langle u_zT\rangle_{A,t}$, 
vertical profiles of mean thermal dissipation $\langle \epsilon_T(z)\rangle_{A,t}$ and mean kinetic energy dissipation $\langle \epsilon(z)\rangle_{A,t}$, 
a $z$-dependent Kolmogorov scale $\langle \eta_K(z)\rangle_{A,t}$ and how well the Gr\"otzbach criterion is satisfied plane by plane $\Delta 
z_e(z)/\langle \eta_K(z)\rangle_{A,t}$. The dashed line in the lower right panel marks $\Delta z_e(z)/\langle\eta_K(z)\rangle_{A,t}=\pi$.
The dashed line in the upper mid panel is the Nusselt number from run FDM3 (see Table \ref{Tab2} or Ref. \cite{Bailon2010}). The inset in the top right
panel magnifies the vertical profile of the mean thermal dissipation rate.}
\label{grid3a}
\end{figure}

\subsection{Statistical properties for resolutions with different polynomial orders}
In correspondence with the so-called h-type and p-type spectral element methods (SEM), two routes of modification of the resolution 
exist. In the h-type  SEM the number of primary elements, $N_e$, is varied, in the p-type SEM  one changes the polynomial degree $N$ 
of the basis functions on each element and keeps the number of elements fixed. In the following, we summarize efforts in both directions 
in order to study resolution effects for the gradient fields. Since the grid is non-uniform in all three directions the side lengths of an element
are functions of the three coordinates, i.e. $\Delta x_e(x,y,z)$, $\Delta y_e(x,y,z)$ and $\Delta z_e(x,y,z)$. Figure \ref{grid_0} (left) shows
a view of the horizontal primary element mesh. The coarsest elements are always found at the cell center line.  
 
Particular emphasis was given here to the vertical resolution since this is the important direction for the correct resolution of the BLs. 
The formula that has been chosen to determine the element boundaries in the vertical direction is given by the following geometric 
scaling for the upper half of the cell with the scaling factor $r$
%---------------------------------------------------------------------
\begin{equation}
\Delta z_e\left(\frac{N_{e,z}}{2}\right)+\dots +\Delta z_e\left(1 \right)=\left(\sum_{k=1}^{\frac{N_{e,z}}{2}}r^{k-1}\right)\Delta z_e\left(1\right)=\frac{1}{2}\,.
\label{geometric}
\end{equation}
%---------------------------------------------------------------------
In correspondence with the up-down-symmetry this relation has to be applied for the lower half as well.  Equidistant vertical meshing 
corresponds to $r=1$. Figure \ref{grid_0} (right) demonstrates the resulting vertical meshing for different numbers of primary element 
nodes, $N_{e,z}$. In the appendix, we describe one way to obtain an optimal non-equidistant grid with respect to $z$, in other words,
an optimal scaling factor $r$ in (\ref{geometric}).
%---------------------------------------------------------------------
\begin{table}
\begin{center}
\begin{tabular}{cccccccccc}
\hline
Run & $N$ & $(N_{e}, N_{e,z})$ & $N_z$ &  $Nu(0)\pm\sigma$ & $Nu(1)\pm\sigma$ & $Nu_V\pm\sigma$ & $\Lambda_T$ & $\Lambda_v$\\
\hline
SEM1 & 3 & (30720, 32) & 96   & 33.2$\pm 0.9$  & 33.2$\pm 0.9$ & 31.5$\pm 2.1$ & 5.6\% & 13.6\%\\
SEM2 & 5 & (30720, 32) & 160 & 31.5$\pm 0.7$  & 31.7$\pm 0.7$ & 31.8$\pm 2.1$ & 0.3\% & 0.6\% \\
SEM3 & 7 & (30720, 32) & 224 & 31.8$\pm 0.9$  & 31.9$\pm 0.9$ & 32.0$\pm 3.1$ & 0.3\% & 0.1\%\\
SEM4 & 11 & (30720, 32) & 352 &  31.6$\pm 0.6$ & 31.9$\pm 0.6$ & 31.8$\pm 2.0$ & 0.1\% & 0.5\%\\
\hline
\end{tabular}  
\end{center}
\caption{Parameters of the different spectral element simulations SEM1 to SEM4. The runs have an identical primary node mesh, 
but different polynomial order on each element.  We display the order $N$ of the Legendre polynomials,
the total number of spectral elements, $N_{e}$, the number of spectral elements with respect to $z$ direction, $N_{e,z}$, 
the number of grid cells resulting from primary and secondary nodes with respect to $z$ direction, $N_z=N_{e,z} N$, and the Nusselt numbers 
$Nu(z=0)$, $Nu(z=1)$ and $Nu_V$. Furthermore we list the relative errors $\Lambda_T$ and $\Lambda_v$ (see Eqns. (\ref{rel1a})).
All four runs are at $Ra=10^8$, $\Gamma=1$ and $Pr=0.7$.}
\label{Tab1}
\end{table}
%---------------------------------------------------------------------

Figure \ref{grid3a} shows important statistical quantities for a variation in correspondence with  a p-type refinement. Results are obtained 
for different polynomial orders but the same primary element mesh. The results are summarized in Table \ref{Tab1}. 
In all cases shown, derivative-based quantities are evaluated spectrally on each element and no derivatives are taken across boundaries. All runs are 
conducted at a Rayleigh number $Ra=10^8$ with a primary element mesh as displayed in Figure \ref{grid_0}. 
On average, we ran our simulations for at least 30 free-fall times $T_f$ to ensure that the system had 
settled into its relaxed state, and then we continued the evolution for at least 75 free-fall times (in case of the biggest DNS), outputting 
on average at least 80 statistically independent snapshots. 

Both the table and the figure indicate that insufficient spectral resolution is  manifested in multiple ways,
but is not necessarily obvious when looking at standard quantities, e.g. the ingredients for the turbulent heat flux. The graphs for the 
mean temperature profile $\langle T(z)\rangle_{A,t}$, the convective heat flux $\sqrt{Ra Pr}\langle u_z T\rangle_{A,t}$ and even the Nusselt 
numbers which are obtained in different ways do not indicate a resolution effect at first glance. However the large magnitude of the relative 
errors of run SEM1 which has been used to test the dimensionless energy balances (\ref{rel1a}) is definitely  caused by the insufficient resolution. 
The  run SEM1 was one of our longest, taking about 1200 $T_f$. The statistical analysis is based on 192 turbulence statistically independent 
three-dimensional snapshots separated by 6 free-fall times each.   
%---------------------------------------------------------------------
\begin{figure}
\centering
\includegraphics[width=0.8\textwidth]{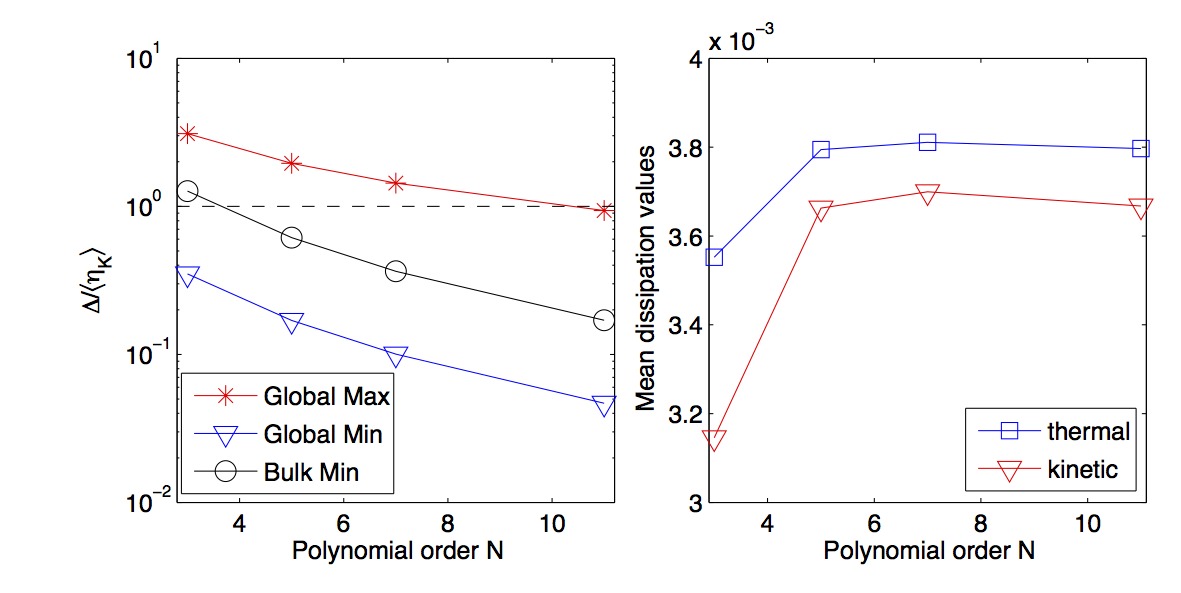}
\caption{Left: Global maximum and minimum grid spacing as well as minimum grid spacing in the bulk region (volume $V_b$) for the 
data in Table \ref{Tab1}.  Data are normalized with respect to mean Kolmogorov length. The definitions are given in (\ref{minmaxgrid1}) 
and (\ref{minmaxgrid2}). Right: Mean dissipation rates of thermal variance, $\langle\epsilon_T\rangle_{V,t}$,  and kinetic energy, 
$\langle\epsilon\rangle_{V,t}$,  as functions of the polynomial order.} 
\label{pdf2}
\end{figure}
%---------------------------------------------------------------------

An increase to $N=5$ in run SEM2 improves the convergence of the energy balances drastically. Nevertheless, the 
plane-averaged mean profiles of the thermal and the kinetic energy dissipation rates still display an insufficient resolution which is present
for the next higher polynomial order, $N=7$, as well. This becomes visible by the discontinuities at the element boundaries, especially near 
the center of the cell. The lower right panel of Fig. \ref{grid3a} relates a refined Kolmogorov-type  scale
%---------------------------------------------------------------------
\begin{equation}
\langle\eta_K(z)\rangle_{A,t}=\frac{Pr^{3/8}}{Ra^{3/8}} \langle\epsilon(z)\rangle_{A,t}^{-1/4}\,.
\label{etaKz}
\end{equation}
%---------------------------------------------------------------------
We can refine the classical Gr\"otzbach criterion (\ref{gbsle1}) to 
%---------------------------------------------------------------------
\begin{equation}
\frac{\Delta z(z)}{\langle\eta_K(z)\rangle_{A,t}}\le \pi\,,
\label{etaKz2}
\end{equation}
%---------------------------------------------------------------------
where $\Delta z$ is the vertical grid spacing, recognizing now element mesh and collocation grid on each element. This is exactly 
what produces the characteristic shape of all the curves in the lower right panel.  
Such a criterion has been suggested already by \cite{Bailon2010}. It shows clearly that all orders $N\le 7$ result in grid spacings 
that are too coarse in the bulk region of the convection cell. In the appendix, we explain how insufficient resolution can cause the 
spike structures in the vertical profiles of the dissipation rates and related quantities by means of a convergence test for a simple analytical profile. 
The artifacts at the element boundaries which we see for the SEM are due to insufficient resolution and hence the failure of the 
derivatives to match at the boundaries, since this SEM method enforces only the continuity of the functions at the boundaries. This gives 
rise to a clear criterion for resolution: when the system is sufficiently resolved, all spikes in both dissipation profiles completely disappear. 
In classical finite difference methods numerical diffusion and dispersion will suppress such spikes. 
In addition it is shown in the appendix that similar observations as in Fig. \ref{grid3a} follow when the h-type route of grid refinement is 
followed, i.e., refining the primary mesh at fixed system parameters $(Ra, Pr, \Gamma)$  and a given polynomial order. To conclude
this part, while some standard indicators for sufficient resolution which have been discussed in previous works \cite{Stevens2010,Shishkina2010} 
are all well-satisfied, a closer look at the dissipation fields indicates clearly that the spatial resolution is not sufficient, in particular in the bulk of the 
cell. The artifacts in the mean vertical profiles of the gradient fields do not completely disappear even when the order is increased to 
$N=11$ as demonstrated in the inset in the top right panel of Fig. \ref{grid3a}.  
 
%---------------------------------------------------------------------
\begin{figure}
\centering
\includegraphics[width=0.8\textwidth]{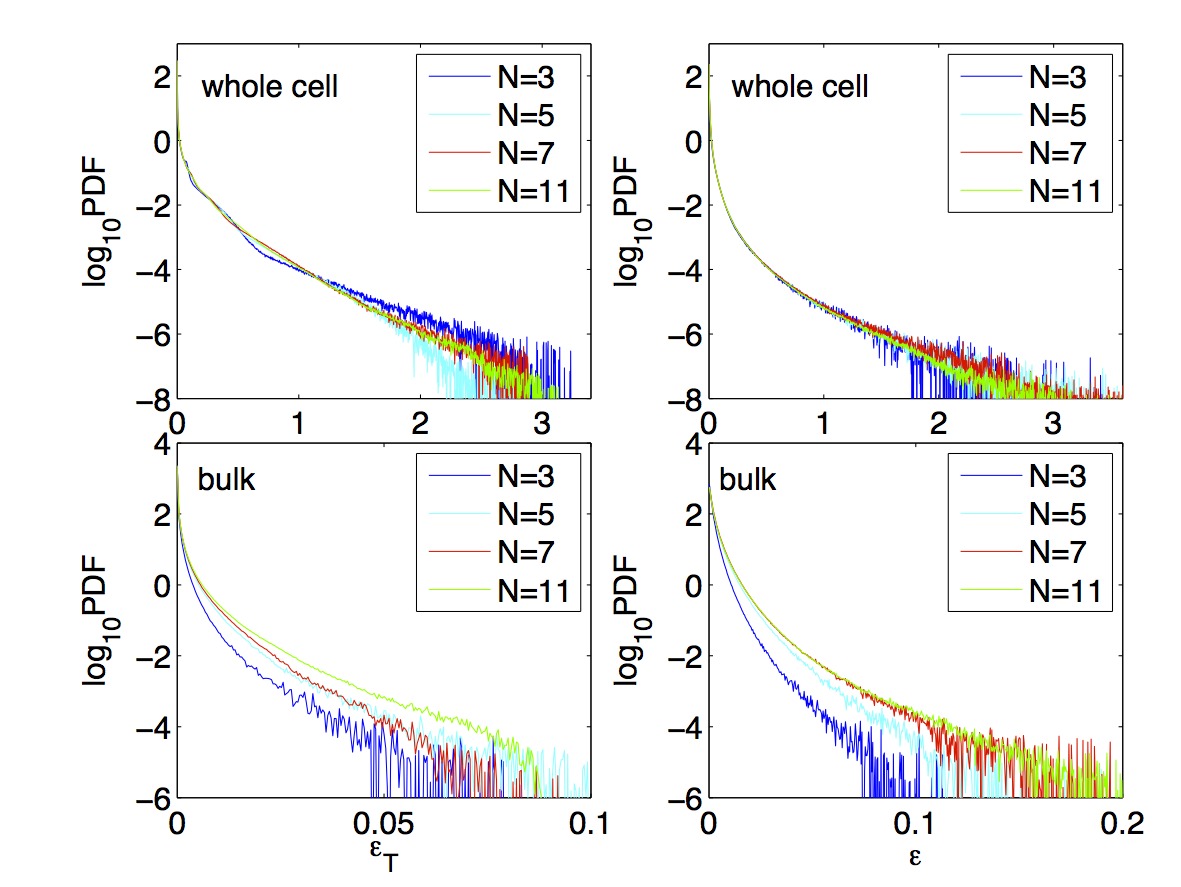}
\caption{Probability density functions (PDFs) of the thermal dissipation rate (left column) and kinetic energy dissipation rate (right column) 
are shown for runs SEM1 to SEM4. The upper row displays the data obtained for the whole cell. The lower row shows the data for the bulk.}
\label{pdf3}
\end{figure}
%-----------------------------------------------------------------

Compared to previous resolution studies of fluid turbulence in periodic boxes \cite{Schumacher2007} and shear flow turbulence channels 
\cite{Hamlington2012}, the situation in the present RB case is more complex. On the one hand, the turbulent flow is inhomogeneous in all space dimensions. 
This causes space-dependent statistical 
properties of the turbulent fields and their derivatives. On the other hand, the computational grid is non-uniform in all three directions as
described already above. Although we refine the grid towards all walls, the regions where one expects the largest amplitude of the derivatives,
it is not necessarily assured that both, steepest gradient and finest grid cells, coincide. In this situation one can however define the coarsest and finest 
grid spacing in the whole cell or a subvolume to get a global indication of the quality of resolution. This is done by the following geometric means
%---------------------------------------------------------------------
\begin{eqnarray}
\label{minmaxgrid1}
\Delta_{max}&=&\sqrt[3]{\max_{x\in {\cal I}}\Delta x_e(x,y,z) \max_{y\in {\cal I}}\Delta y_e(x,y,z) \max_{z\in {\cal I}_z}\Delta z_e(x,y,z)}\,\\
\Delta_{min}&=&\sqrt[3]{\min_{x\in {\cal I}}\Delta x_e(x,y,z) \min_{y\in {\cal I}}\Delta y_e(x,y,z) \min_{z\in {\cal I}_z}\Delta z_e(x,y,z)}\,
\label{minmaxgrid2}
\end{eqnarray}
%---------------------------------------------------------------------
with ${\cal I}=[-0.5,0.5]$ or a subinterval and ${\cal I}_z=[0,1]$ or a subinterval, such as the bulk volume $V_b$ which is given further below in the
text. In Fig. \ref{pdf2} (left) we display the minimum and maximum grid spacing for the whole cell obtained by (\ref{minmaxgrid1}) and (\ref{minmaxgrid2}).
Furthermore, we show the minimum resolution in the bulk region where we defined a subvolume $V_b=\{ {\bf x}=(r,\phi,z)\;|\, 0\le r\le 0.3; 0.2\le z\le 0.8\}$. 
The right panel confirms what we have discussed already above, that the mean dissipation rates level off for $N\ge 5$ although vertical profiles are still
not sufficiently well resolved.
%---------------------------------------------------------------------
\begin{figure}
\centering
\includegraphics[width=1.0\textwidth]{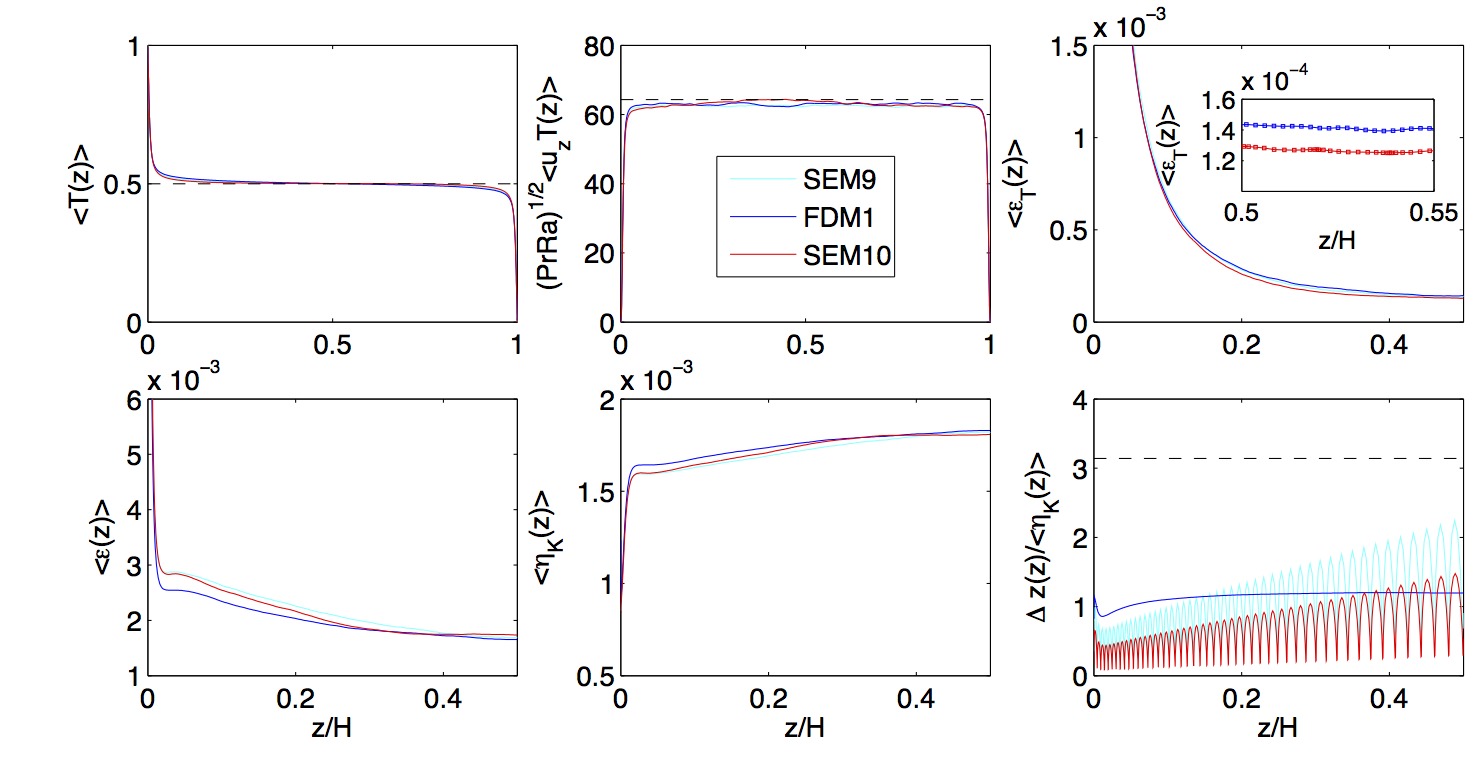}
\caption{Comparison of spectral element  runs SEM9 and SEM10 with finite difference run FDM1 (see Table \ref{Tab2} for specifications).  The same 
quantities are plotted as in Figure \ref{grid3a}. The dashed line in the lower right panel marks $\Delta z(z)/\langle\eta_K(z)\rangle_{A,t}=\pi$.
The dashed line in the upper mid panel is the Nusselt number of the FDM run from \cite{Bailon2010}) with $Nu=64.3$. The inset in the top right panel magnifies
the thermal dissipation rate profiles for FDM1 and SEM10.}
\label{compare}
\end{figure}
%---------------------------------------------------------------------

In Figure \ref{pdf3} we display the probability density functions (PDFs) of the fields $\epsilon_T$ and $\epsilon$. 
The upper row shows data which have been obtained in the whole cell, the lower row those for the subvolume $V_b$ in the bulk. It can be 
seen for all four panels that with increasing polynomial order more very-high-amplitude events are resolved and that the tail is further stretched out.
The better resolution manifests in significantly less scatter at the largest amplitudes. Even more pronounced are the resolution effects in the bulk 
(lower row). We observe now for both dissipation rates the same systematic trend. The tail of the stretched exponential 
distribution is fatter for higher polynomial order.  This latter finding is also in agreement with previously reported spectral resolution studies for 
homogeneous isotropic box turbulence as reported in \cite{Schumacher2007}. Note that the tails of the exponents in our figure do not 
always increase in an even manner with resolution. One sees for example, a jump in the lower left panel of Figure \ref{pdf3} in going from N=3 to 
N=5 and then again from N=7 to N=11. This is understandable in light of Figure \ref{extreme}, where we see that high-amplitude events can increase 
the tails significantly. Since the system is chaotic as well as turbulent, simulations done for the same Rayleigh number but different resolution are 
statistically different, so some of them could have more high amplitude events than others. Longer simulation times would help smooth this out.

%-----------------------------------------------------------------
\begin{table}
\begin{center}
\begin{tabular}{cccccccccc}
\hline
Run  & $Ra$ &  $Pr$ &$N_{e}$ & $N_{e,z}$ & $N_z$ & $N_eN^3$ & $Nu_V$ & $\Lambda_T$ & $\Lambda_v$\\
\hline
SEM5 & $10^6$  & 0.7 & 30720 & 32 & 224 & $1.05\times 10^7$ & 8.6 & 0.1\% & 0.1\%\\
SEM6 & $5\times 10^6$ & 0.7 & 30720 & 32 & 224 & $1.05\times 10^7$ & 13.9 & 0.3\%& 0.5\%\\
SEM7 & $10^7$   & 0.7 & 30720 & 32 & 352 & $4.08\times 10^7$ & 16.6 & 0.3\% & 0.6\%\\
SEM7a & $10^7$ &6.0 & 30720 & 32 & 352 & $4.08\times 10^7$ & 16.6  & 0.7\%& 0.2\%\\
SEM8 & $10^8$  & 0.7 &  256000 & 64 & 704 & $3.41\times 10^8$ & 31.4 & 0.4\% & 0.2\%\\
SEM9 & $10^9$  &  0.7 & 875520 & 96 & 672 & $3.00\times 10^8$& 62.8 & 0.1\%& 0.1\%\\ 
SEM10 & $10^9$  & 0.7 &  875520 & 96 & 1056 & $1.17\times 10^9$& 63.1 & 1.2\%& 0.5\%\\ 
FDM1 & $10^9$  & 0.7 & -- & -- & 621 &$1.62\times 10^8$& 63.1 & 1.7\%& 9.2\%\\ 
\hline
\end{tabular}  
\end{center}
\caption{Parameters of the different spectral element simulations. Runs SEM7, SEM8 and SEM10 have an order of the 
Legendre polynomials $N=11$, runs SEM5, SEM6 and SEM9 use $N=7$. 
We show the  Rayleigh number $Ra$, the Prandtl number $Pr$, 
the total number of spectral elements, $N_{e}$, the number of spectral elements with respect to $z$ direction, $N_{e,z}$, 
the number of grid planes (primary and secondary nodes) with respect to $z$ direction, $N_z$, the total number of grid cells $N_eN^3$,  
and the Nusselt number $Nu_V$. Furthermore we list the relative errors $\Lambda_T$ and $\Lambda_v$ (see Eqns. (\ref{rel1a})).
All simulations are conducted for $\Gamma=1$ and $Pr=0.7$. The finite difference run FDM1 is conducted for
$N_{\phi}\times N_r\times N_z=721\times 361\times 621$, respectively.}
\label{Tab2}
\end{table}
%---------------------------------------------------------------------
\subsection{Comparison with the finite difference method}
The comparison of DNS runs at $Ra=10^9$ and $Pr=0.7$, i.e. runs SEM9, SEM10 and FDM1 from Table \ref{Tab2}, is displayed in Figure \ref{compare}. 
The resolution of FDM1 has been chosen twice as fine as in the run from Bailon-Cuba et al.
\cite{Bailon2010} in order to get a comparable number of grid points with respect to SEM9. The thermal boundary layer for FDM1 is resolved with 21 grid planes.  We see that the agreement of the mean vertical profiles is very good. 
The value of the Nusselt numbers for the higher resolution runs match now to within three significant figures, also with the data from Ref. \cite{Wagner2012}. In Table
\ref{Tab2}, we list also the corresponding relative errors $\Lambda_T$ and $\Lambda_v$ which are larger for FDM1 than those for  SEM9 and SEM10,  particularly 
for the kinetic energy dissipation rate in our study. The latter has been evaluated in correspondence with (\ref{kinetic}) which has to be applied for inhomogeneous
flows. The difference is supported by the deviation in the vertical profiles of $\langle \epsilon(z)\rangle_{A,t}$ in the lower left panel of  Fig. \ref{compare}.  We also 
note that Stevens et al. (see their Table 1 in \cite{Stevens2010}) reported similar errors which were in their case however mostly detected for the thermal energy dissipation 
rate and not only for the lowest resolutions.

\begin{figure}
\centering
\includegraphics[width=0.95\textwidth]{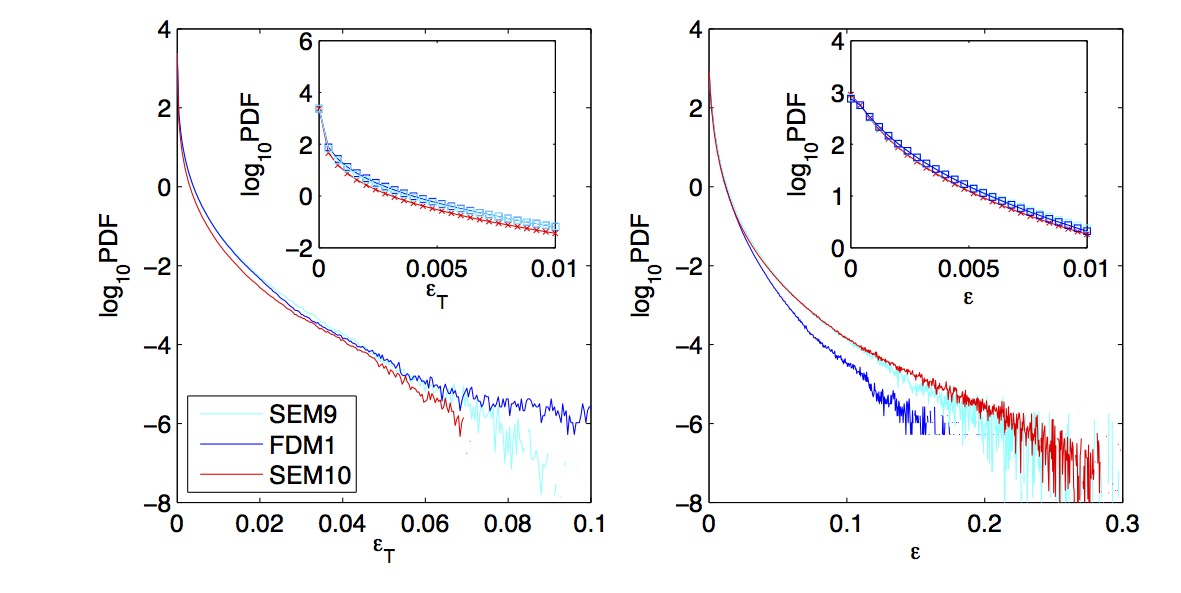}
\caption{Comparison of the PDFs of $\epsilon_T$ (left) and $\epsilon$ (right) obtained in the subvolume $V_b$.
The insets magnify the smaller amplitudes. We compare the data for FDM1 
with those from SEM9 and SEM10. Line colors are the same for both figures and indicated in the legend and agree with those 
from Fig. \ref{compare}.}
\label{comppdf}
\end{figure}

The distribution of the local amplitudes of both dissipation fields is compared in Fig. \ref{comppdf}. Both panels show that the deviations arise mostly 
for the outer tails where the extreme fluctuations are captured. In case of the thermal dissipation 
rate both PDFs remain closely together for nearly the whole range. The kinetic energy dissipation rate data start to differ for roughly 15--20\% of the maximum
amplitude which might be one reason for the larger values of $\Lambda_v$. The agreement in the low-amplitude part
of the PDFs is good as shown in both insets. 

\begin{figure}
\centering
\includegraphics[width=1.0\textwidth]{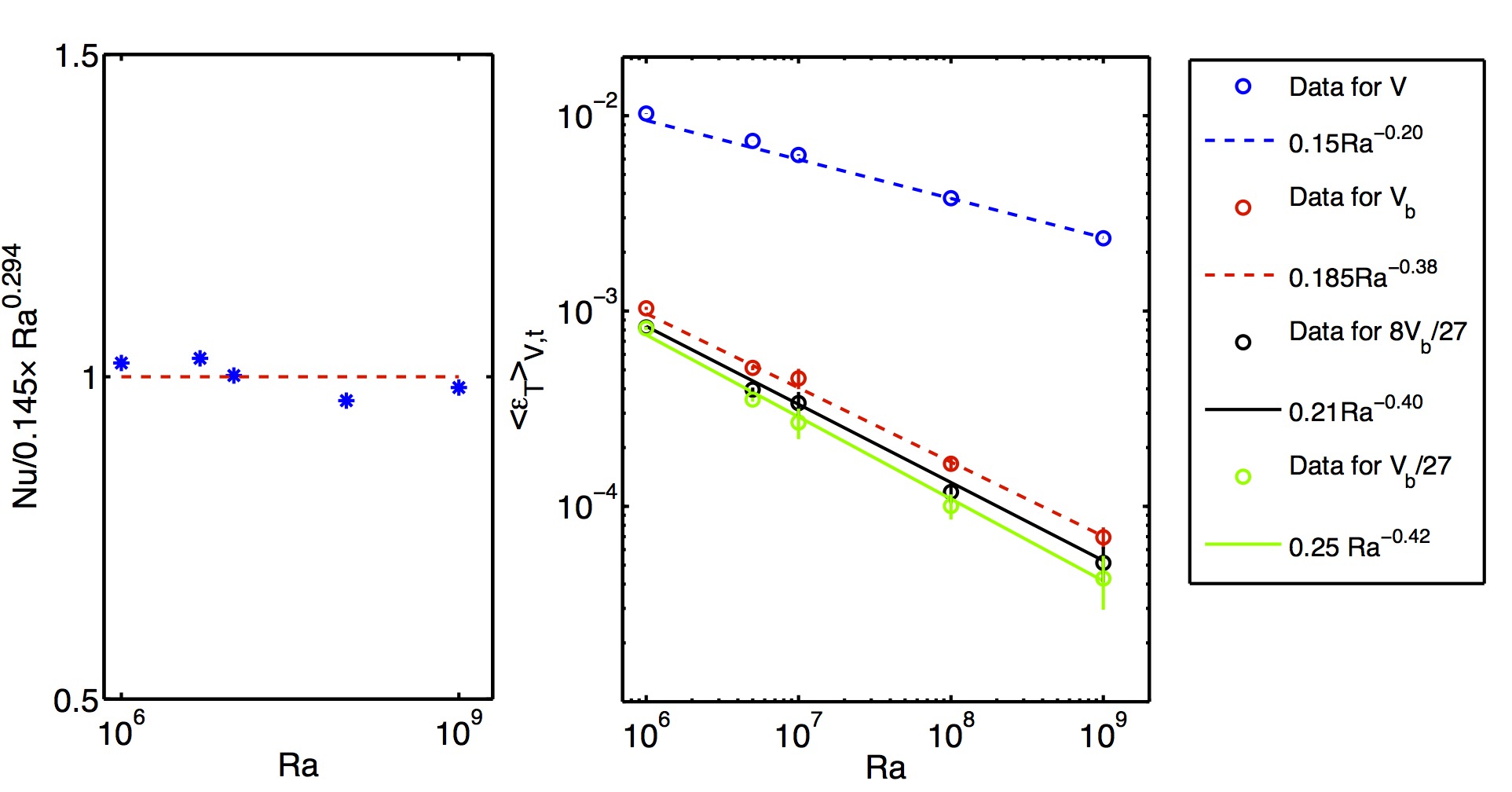}
\caption{Comparison of global transport properties between SEM and FDM. Left: Nusselt versus Rayleigh number for
$10^6 < Ra < 10^9$. The data are compensated by the power law $Nu=0.145 \times Ra^{0.294}$ which was a fit to the data as reported
in Bailon-Cuba et al. \cite{Bailon2010}.  Right: mean thermal dissipation as a function of the Rayleigh number for different subvolumes. The first two data sets are compared with fits to the FDM data (shown as dashed lines). For the 
whole cell $V$ we take former results from Ref. \cite{Emran2012}, for the bulk volume $V_b$ we  compare with data from Ref. \cite{Emran2008}.  
In this case the prefactor is different since the subvolume $V_b$ was chosen differently. The last two series are obtained in smaller subvolumes 
and are fitted by power laws as given in the legend and shown as solid lines.}
\label{comptrans}
\end{figure}

Furthermore, we find excellent agreement between the two codes when comparing global transport properties. For example, the Nusselt 
number and the globally averaged thermal dissipation rates for both the whole cell and the bulk volume $V_b$ as shown in Fig. \ref{comptrans}
agree quite well. We varied our Rayleigh number between $10^6$ and $10^9$ and compared with fits of the FDM data from \cite{Bailon2010}, 
\cite{Emran2012} and \cite{Emran2008}, respectively.  We did need to use a different prefactor for the bulk-averaged thermal dissipation rate 
since our subvolume $V_b$ was chosen differently.

To estimate the effect of the size of our subvolume on the thermal dissipation rates, we show two addtional data sets in the right panel of Fig. 
\ref{comptrans}: one for a smaller subvolume $8V_b/27$ and the other for an even smaller subvolume of $V_b/27$ (all centered about the middle of the cell).  
The general trend is for the thermal dissipation rates to slightly decrease as the subvolume decreases. Also, our uncertainty becomes larger as 
the subvolume decreases. We estimated the uncertainty in the mean $\langle\epsilon_T\rangle_{V,t}$ values  by computing the difference between 
the mean taken over the entire time series and the mean taken over only the latter half of the time series.

The fits to the  data sets corresponding to the smallest subvolumes are $\langle\epsilon_T\rangle_{8V_b/27,t} = (0.21\pm 0.07)Ra^{-0.40\pm 0.02}$  
and $\langle\epsilon_T\rangle_{V_b/27,t} = (0.25\pm 0.12)Ra^{-0.42\pm 0.03}$. Kaczorowski and Xia \cite{Kaczorowski2013} also studied the 
scaling of subvolume-averaged thermal dissipation rates in a similar range of Rayleigh numbers using a  small subvolume ($V/64$) but for a 
Prandtl number of 4.38. Our exponent disagrees with theirs of $\langle\epsilon_T\rangle_{V,t} =  43.9 Ra^{-0.78}$. We do see a trend towards a 
larger exponent as our subvolume decreases,  but our largest exponent still disagrees with  \cite{Kaczorowski2013} even when including our 
estimates of numerical uncertainty.

%-----------------------------------------------------------------
%-----------------------------------------------------------------
%---------------------------------------------------------------------
\begin{figure}
\centering
\includegraphics[width=0.48\textwidth]{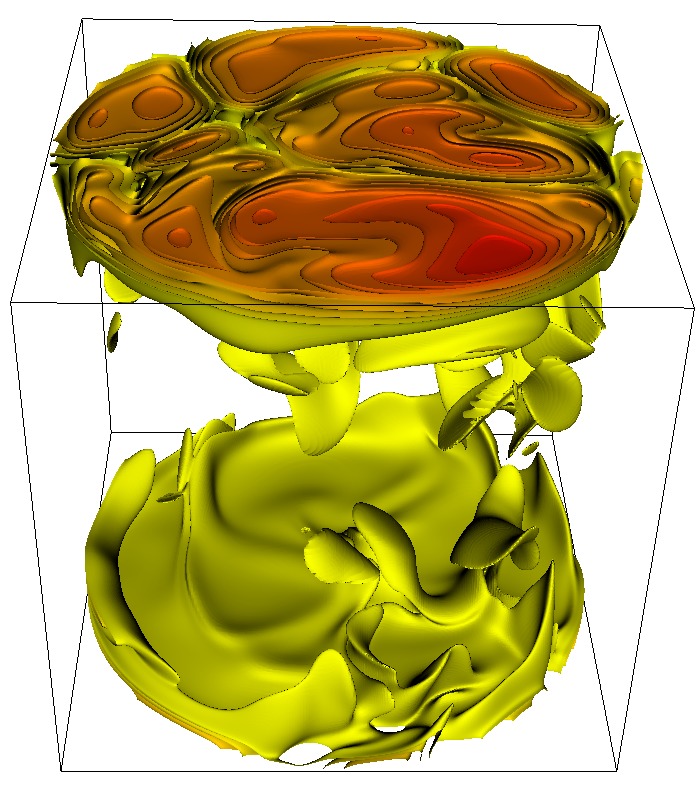}
\includegraphics[width=0.48\textwidth]{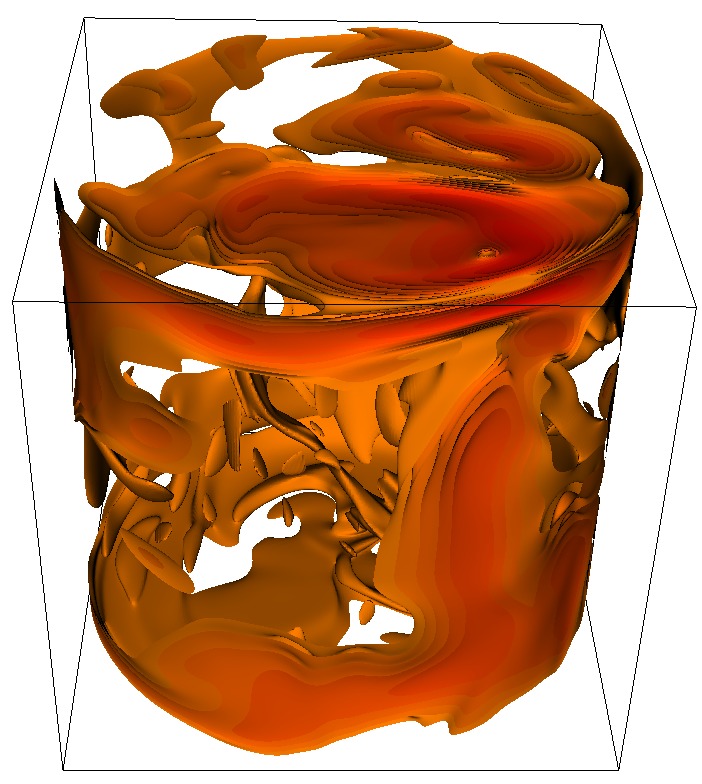}
\caption{Isovolume plots for the thermal dissipation rate (left) and the kinetic energy dissipation rate (right).
Data are obtained for $Ra=10^7, Pr = 0.7$ and are shown in logarithmic units. Left: The range of the data is
$\log(\epsilon_T)\in [-21.1, 0.5]$ and isosurfaces between -5 and 0 are shown. Right: The range of the data is
$\log(\epsilon)\in [-37, 0.0]$ and isosurfaces between -4.5 and 0 are shown. Data are from SEM7 in Table 2.}
\label{highamp}
\end{figure}
%---------------------------------------------------------------------
\begin{figure}
\centering
\includegraphics[width=1.0\textwidth]{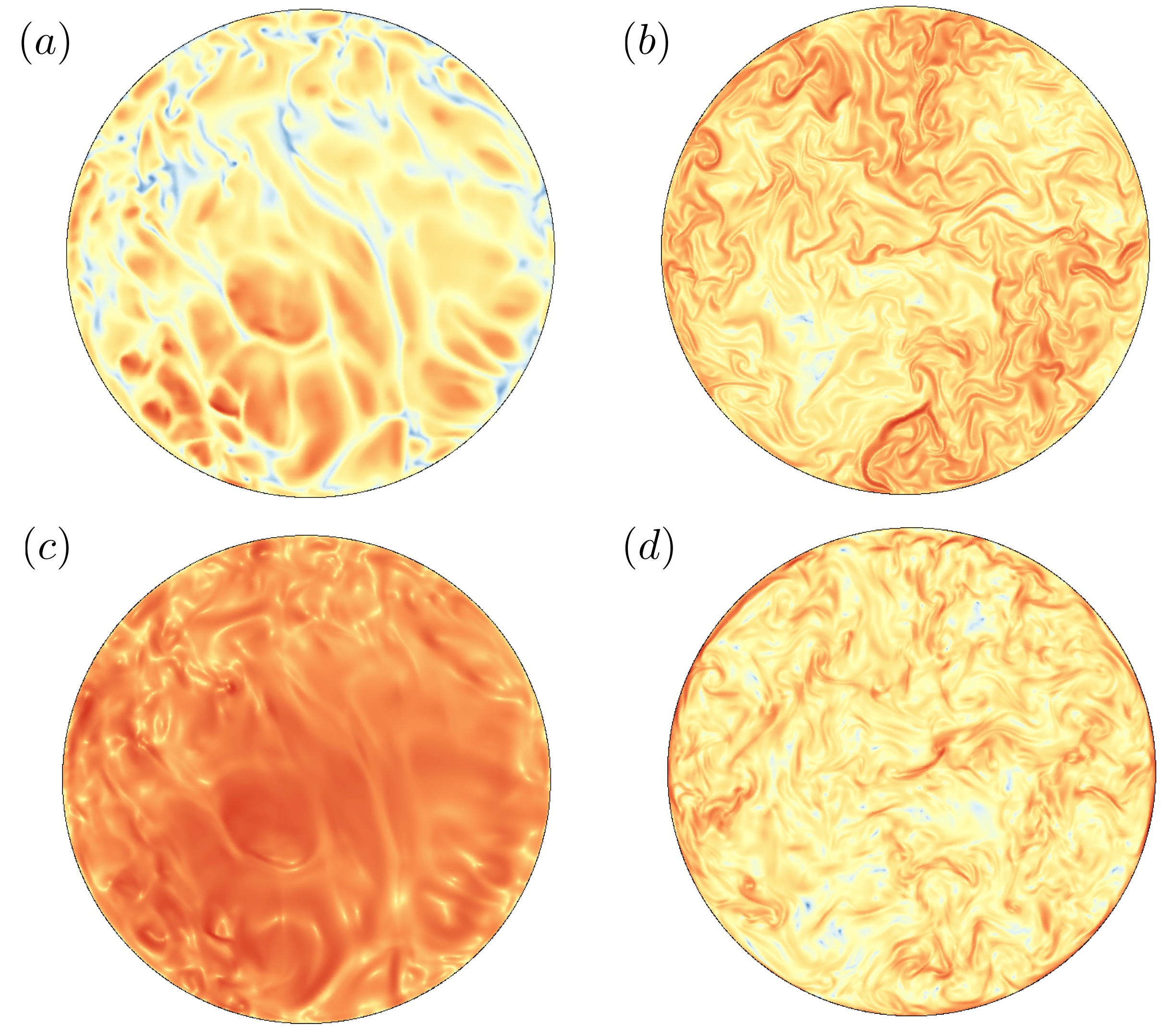}
\caption{Snapshots of the local thermal dissipation rate $\epsilon_T(x,y)$ for (a) $z=0$ and (b) $z=0.5$ and the 
local kinetic energy dissipation rate $\epsilon(x,y)$ also for (c) $z=0$ and (d) $z=0.5$. Data are obtained for 
$Ra=10^9, \Gamma = 1.0, Pr = 0.7$ and are shown in logarithmic units. Data are from SEM10 in Table 2. The range of the data is 
$\log(\epsilon_T)\in$ (a) $[-2.8,0.3]$, (b) $[-11.0, -2.0]$, and $\log(\epsilon)\in$ (c) $[-10.0,0.0]$, (d) $[-6.0,-1.2]$.}
\label{epstvslices}
\end{figure}
%-----------------------------------------------------------------

\section{Results}
\subsection{Very-high-resolution runs at different Rayleigh numbers}

In the following section, we want to discuss a series of very-high-resolution runs in more detail. All the runs with their 
resolution are displayed in Table \ref{Tab2}. We first compare runs at $Pr=0.7$ spanning a Rayleigh number range from
$10^7$ to $10^9$. 

Snapshots of high-amplitude regions of both dissipation fields are shown in Figure \ref{highamp}. The data are given in 
logarithmic units. Both dissipation rates form smooth sheet-like structures in the bulk, in particular the thermal dissipation rate. 
The very fine resolution is clearly obvious from the absence of ripples at the isosurfaces of both dissipation fields. 
In Figure \ref{epstvslices} we show horizontal slices of both dissipation fields at fixed height $z$. Again the data is given in 
logarithmic units to highlight the variation. The top row is for $\epsilon_T(x,y)$ and the bottom corresponds to $\epsilon(x,y)$. 
The left column is for the bottom plate, $z=0$, and the right column is for the midplane, $z=0.5$. We  see a smaller range of 
scales at midplane than near the bottom plate, consistent with Figure \ref{pdf3}. The fine filamentary structure present in this case is
similar to passive scalar turbulence \cite{Watanabe2004} or convectively driven mixing layers \cite{Mellado2010}. Interestingly the thermal dissipation rate 
appears to be correlated with the kinetic dissipation rate at the bottom plate, but less so at the midplane. At the bottom plate the structures
reflect the ongoing plume formation and detachment.

\begin{figure}
\centering
\includegraphics[width=0.8\textwidth]{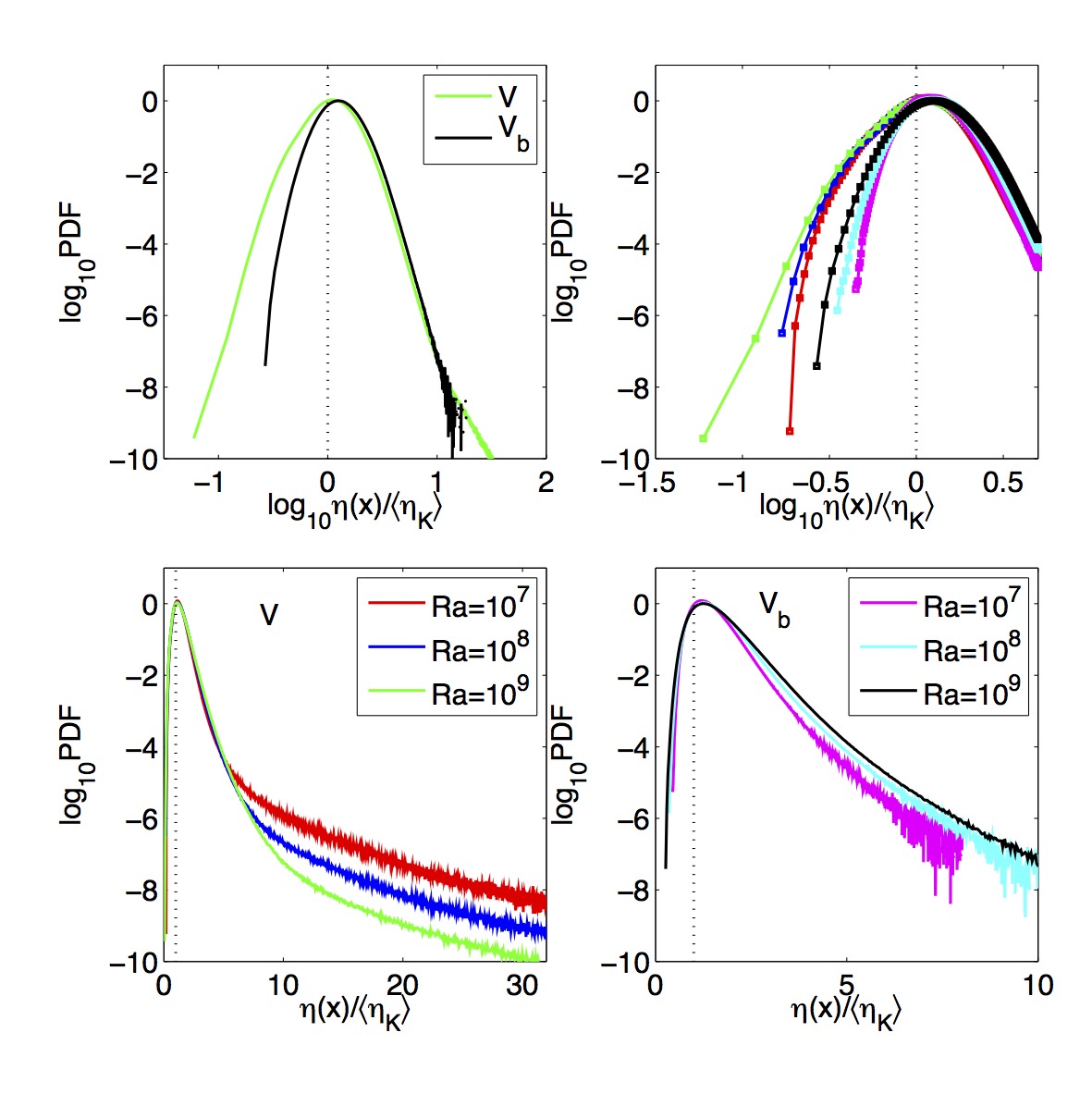}
\caption{Probability density function (PDF) of the local dissipation scale $\eta({\bf x}, t)$ for runs SEM7, SEM8 and SEM10 as given in 
the legend and Table \ref{Tab2}. We compare the PDFs obtained in the whole convection cell with volume $V$ and  those obtained in the bulk 
which is defined as the subvolume $V_b$. The dotted lines indicate $\eta=\langle\eta_K\rangle$. Top left: Comparison of the PDFs in $V$ and $V_b$ for
$Ra=10^9$. Top right: Magnification of the left tails for all six data sets. Bottom left: Rayleigh number dependence in the whole cell. Bottom right: 
Rayleigh number dependence in the bulk. Line colors are uniquely chosen for all plots and indicated in the legends.}
\label{pdf4}
\end{figure}

\begin{figure}
\centering
\includegraphics[width=0.8\textwidth]{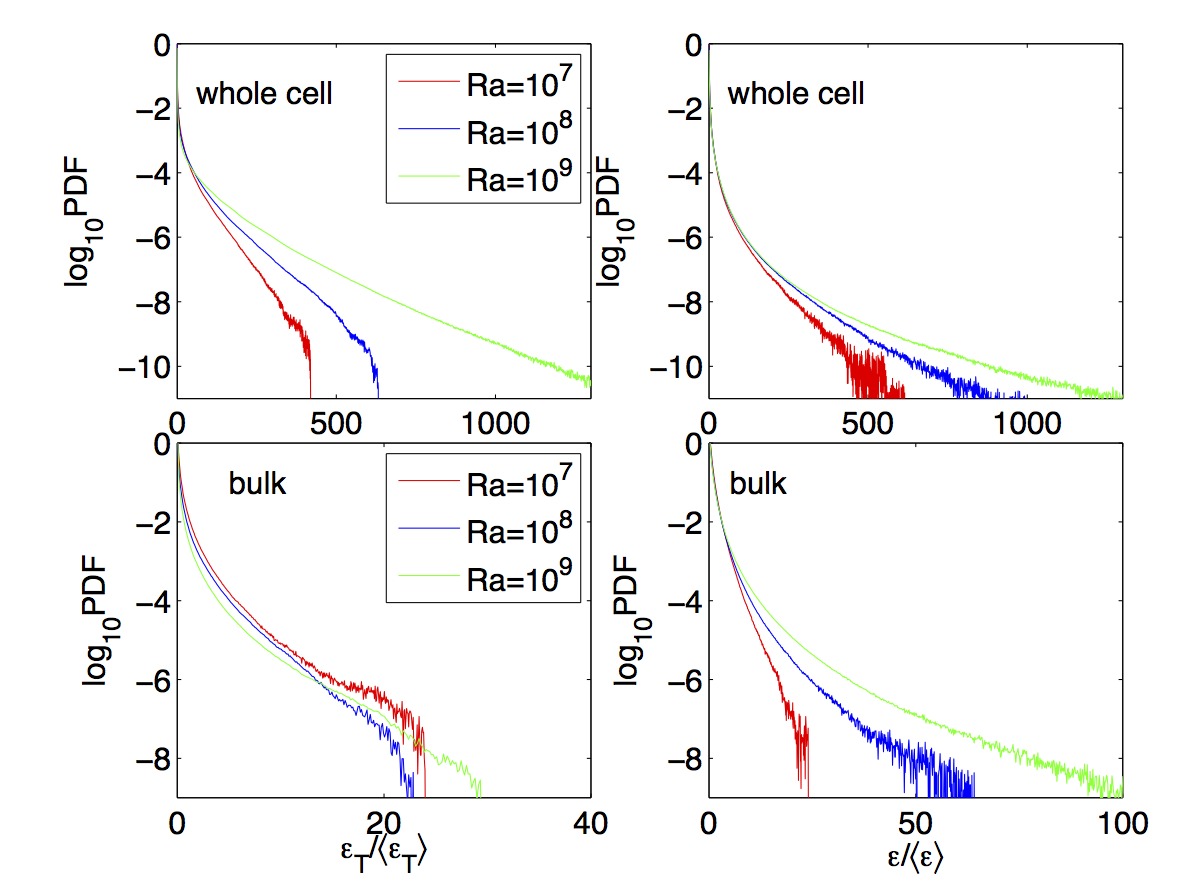}
\caption{Probability density function (PDF) of the dissipation rates for runs SEM7, SEM8 and SEM10 as given in 
the legend and Table \ref{Tab2}. We compare the PDFs obtained in the whole convection cell (upper panels) and  those obtained in the bulk 
which is defined as the subvolume $V_b$ (lower panels). Thermal dissipation rates are displayed
in the left column, energy dissipation rates in the right one. }
\label{pdf5}
\end{figure}

The distribution of the locally fluctuating dissipation scales $\eta_K({\bf x})$ as defined in (\ref{diss1a}) is shown Fig. \ref{pdf4} 
for runs SEM7, SEM8 and SEM10. The scales have been analyzed in the whole cell with volume $V$ as well as in a bulk region which is defined by $V_b$. The 
definition (\ref{diss1a}) has been chosen for this analysis which can be straightforwardly applied to the non-uniform grids that have been 
used for all DNS. An alternative definition of local dissipation scales which is based on velocity increments  was suggested in 
\cite{Yakhot2005,Yakhot2006}. In Ref. \cite{Hamlington2012}, it was shown how both distributions can be related to each other. It can be observed first 
that the scales in the whole cell cover a wider range, both, to the large- and small-scale end (see top left panel) which is centered around the most probable value 
which is always close to mean dissipation scale $\langle\eta_K\rangle_{V,t}$  which is calculated following (\ref{diss1b}). This finding
is also in agreement with previous DNS results \cite{Emran2008,Emran2012} which show that dissipation rates have significantly higher amplitudes 
in the boundary layers. We also see that the right tail ends of the distributions in the whole cell decrease with Rayleigh number. It 
demonstrates that the scales in turbulent RBC become finer as the Rayleigh number increases. This argument is also supported by the fact that the 
differences between the distributions in $V$ and $V_b$ become smaller. In the top right figure, we zoom into the left tail end for all six data sets. The 
smallest local dissipation scales are associated with the largest dissipation events which arise for very steep local gradients. With increasing 
Rayleigh number these contributions become larger, i.e. the left tail becomes fatter. A similar behavior, however much less 
pronounced, can be observed if one restricts the analysis to the bulk volume $V_b$. 
%---------------------------------------------------------------------
%---------------------------------------------------------------------
In Figure \ref{pdf5}, we display the PDFs of both dissipation rates in the whole cell and in the bulk. For both rates, it can be clearly seen that the major 
contribution to the high-amplitude events comes from the boundary layer regions. This has been studied already in \cite{Emran2008}. 
%---------------------------------------------------------------------
%---------------------------------------------------------------------
\begin{figure}
\centering
\includegraphics[width=0.8\textwidth]{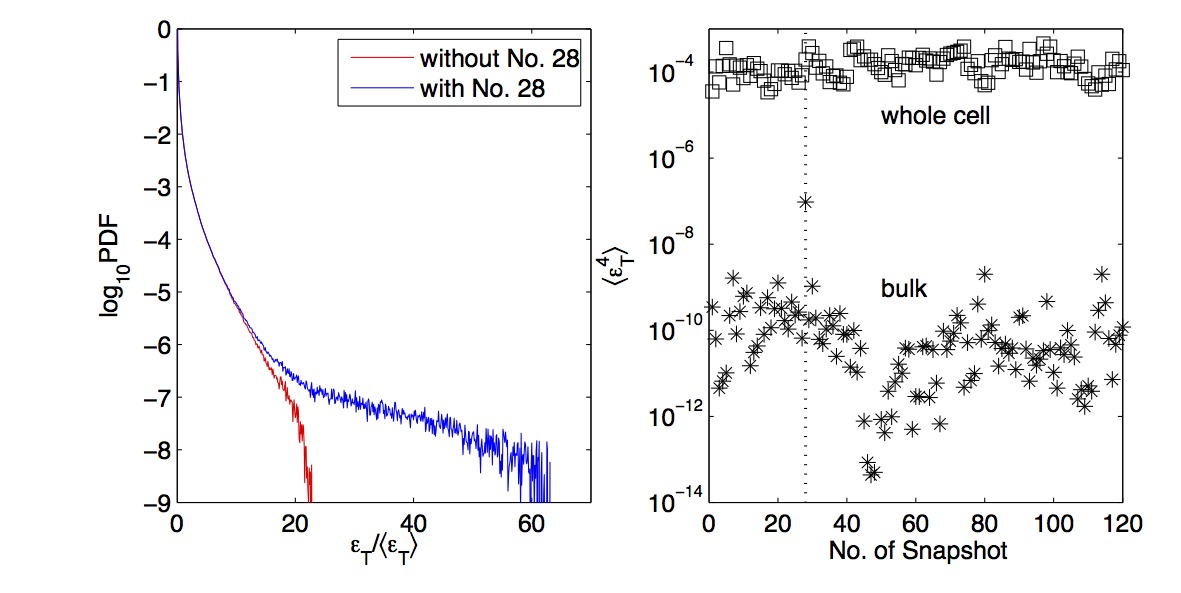}
\caption{Impact of high-amplitude dissipation events on the tail of the probability density function (PDF) of the thermal dissipation rate for SEM8.  
Left: PDFs with and without the high-amplitude event (snapshot No. 28). Right: Fourth order dissipation rate moments in the whole cell and the bulk. 
The dashed line indicates snapshot No. 28.}
\label{extreme}
\end{figure}
%---------------------------------------------------------------------
\begin{figure}
\centering
\includegraphics[width=0.95\textwidth]{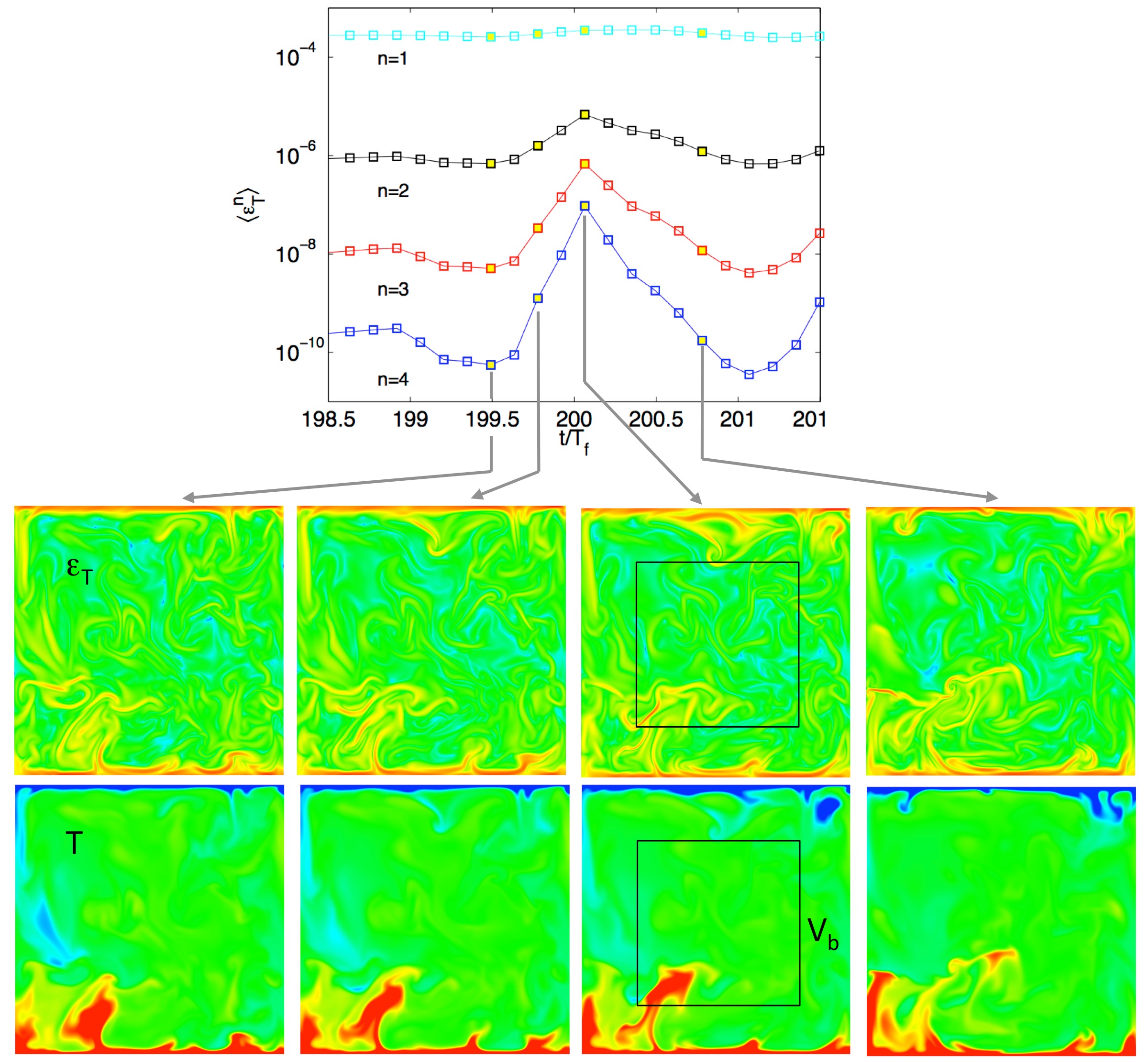}
\caption{The dynamical sequence of the high-amplitude dissipation event from Fig. \ref{extreme}. The top panel displays the temporal evolution of 
the thermal dissipation rate moments of order $n=1$ to 4 calculated in $V_b$. The times which correspond to the images have been plotted as symbols filled with yellow. The first row 
below the top panel displays vertical slice images of the logarithm of the thermal dissipation rate field taken in the plane $(x=0,y,z)$. The range of the data is $\log(\epsilon_T) \in [-20,0.1]$. The bottom row shows the corresponding 
temperature and the range of the data is $T \in [0.3, 0.7]$. We also highlight the size of the subvolume $V_b$ in comparison to the full volume $V$. Data are obtained for run SEM8.}
\label{extreme1}
\end{figure}
%---------------------------------------------------------------------

Figure \ref{extreme} illustrates the sensitivity of the statistics with respect to a single extreme event that was monitored in the course of the 
simulation run and can be identified as a large scale plume sweeping through the bulk volume. It causes a large instantaneous thermal dissipation
which is not easily detectable in the mean dissipation $\langle\epsilon_T(t)\rangle_{V_b}$. Only in the fourth moment of the thermal dissipation $\langle\epsilon_T^4\rangle$ which is 
taken in the bulk volume does this strong event become clearly visible as seen in the right panel of figure \ref{extreme}. Note also that the fourth moment in the 
whole cell is also fairly insensitive to this particular high-amplitude bulk event. The impact of this single event on the statistics is shown in the left panel 
of Figure \ref{extreme}. As expected the tail is stretched significantly. 

In Fig. \ref{extreme1} we display a time sequence of the dynamics which is associated with this single extreme event. The upper panel of the figure
replots the moments of the thermal dissipation rate obtained in $V_b$ with respect to time but on a finer time scale. It can be observed that the entire event lasts less than two free-fall times. We find that the fourth-order moment increases by about three orders of magnitude within this short period. 
Figures \ref{extreme} and \ref{extreme1}  clearly show that the statistical fingerprint of this strong event is best detected in the higher-order moments. The bottom panels of Figure \ref{extreme1} show vertical slice images of the thermal dissipation rate and the temperature corresponding to four different times in the evolution of this event.  Clearly visible is the pronounced hot plume rising and then detaching from the bottom plate which generates steep temperature gradients and thus a large amplitude of the thermal dissipation rate.
%---------------------------------------------------------------------
\begin{figure}
\centering
\includegraphics[width=0.7\textwidth]{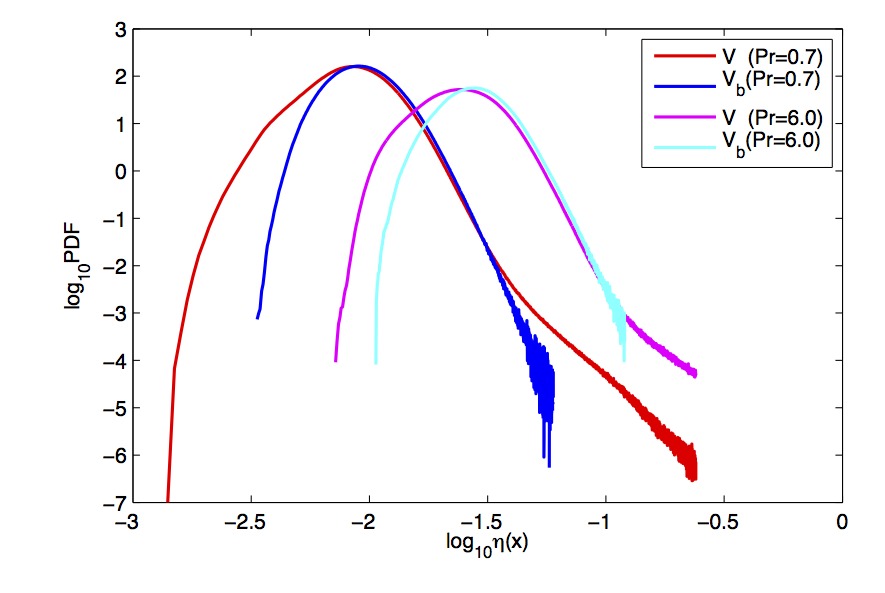}
\caption{Probability density functions of the local dissipation scales obtained in the whole cell and the bulk subvolume for SEM7 and SEM7a.}
\label{etak2}
\end{figure}
%---------------------------------------------------------------------

\subsection{Very-high-resolution run at higher Prandtl number}
Lastly, we compare the gradient statistics at a given Rayleigh number for two different Prandtl numbers. Runs SEM7 and SEM7a
are conducted at $Ra=10^7$ and $Pr= 0.7$ (air) and 6 (water). The data in Table  \ref{Tab2} indicates already that the resolution requirements  remain 
the same for the enhancement of the Prandtl number by a factor of nearly 10. For $Pr>1$, the mean diffusion scale of the temperature field, 
$\langle\eta_B\rangle$ is smaller than mean Kolmogorov scale, $\langle\eta_K\rangle$, since a viscous-convective range on scales smaller 
than the Kolmogorov scale builds up.  Figure \ref{etak2} displays the distributions of $\eta_k({\bf x},t)$. We observe again that the range of varying
scales is larger when the data are taken in the whole cell in comparison to the bulk volume.

\begin{figure}
\centering
\includegraphics[width=0.8\textwidth]{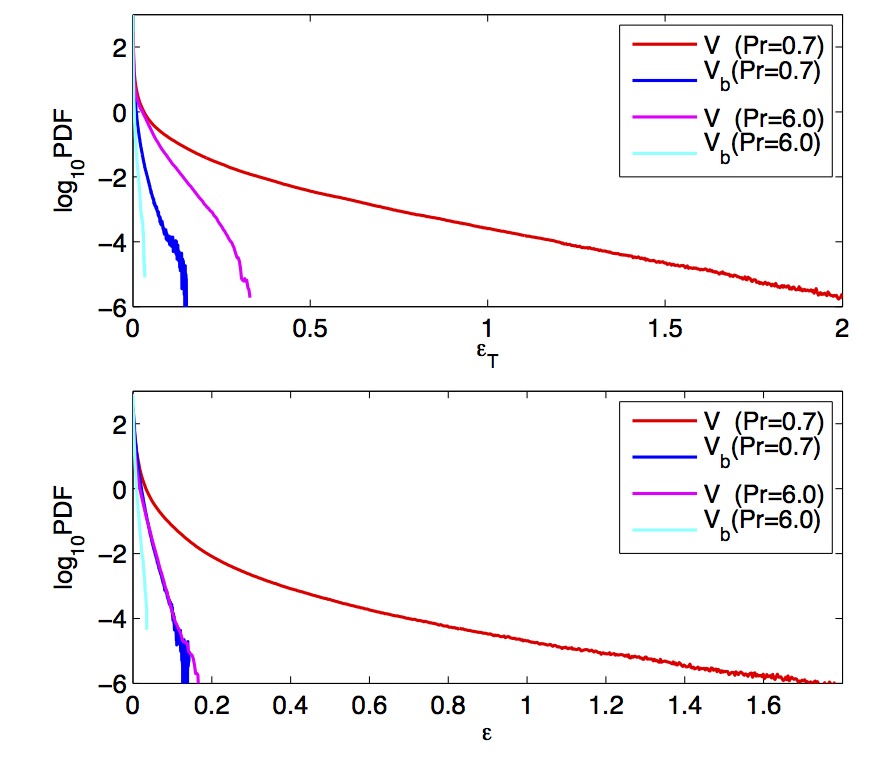}
\caption{Probability density functions of the thermal dissipation (top) and kinetic energy dissipation rates (bottom), respectively. We compare 
runs SEM7 and SEM7a for the analysis in the bulk $V_b$ and the whole cell $V$.}
\label{pdfPrandtl}
\end{figure}

We have not rescaled the distributions by the corresponding
mean dissipation scale since we want to point to the shift of both PDFs for $Pr=6$. This means that the local dissipation scales are larger as a whole 
than for the case of $Pr=0.7$.  This behavior looks counter-intuitive at first glance, particularly from the perspective of passive scalar mixing 
at increasing Prandtl (or Schmidt) number \cite{Schumacher2005}. There one detects increasingly finer diffusion scales for the passive scalar leaving
however the local dissipation scales unchanged. For turbulent convection, we estimated already in the introductory part that the relatively slow falloff 
of $\langle\eta_B\rangle\sim Pr^{-1/8}$ 
as we progress from $Pr\approx 1$ to $Pr\gg 1$. It is the active nature of the temperature field which causes this different behavior in the convection case
 as compared to passive mixing. A temperature field at a higher Prandtl number exists on finer scales than a velocity field, obeying narrower plume structures which  causes a weaker driving of the fluid motion resulting in less steep velocity gradients and consequently larger local dissipation scales. This argumentation is also
supported by a comparison of the PDFs of both dissipation fields as shown in Fig. \ref{pdfPrandtl}. High-amplitude events and tails are shifted to smaller magnitudes
in all cases.
%---------------------------------------------------------------------
%---------------------------------------------------------------------
\begin{figure}
\centering
\includegraphics[width=0.49\textwidth]{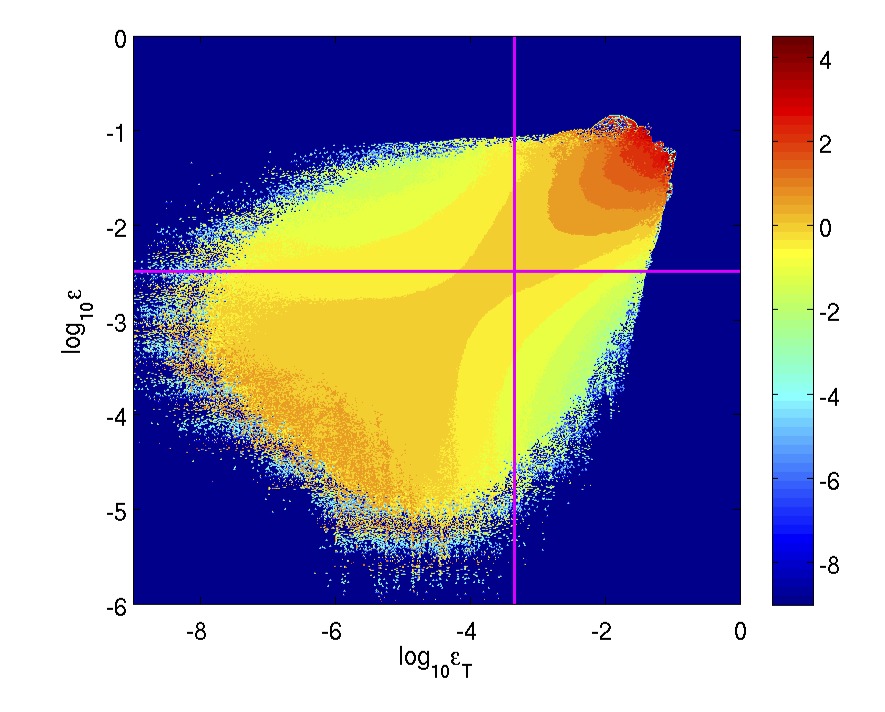}
\includegraphics[width=0.49\textwidth]{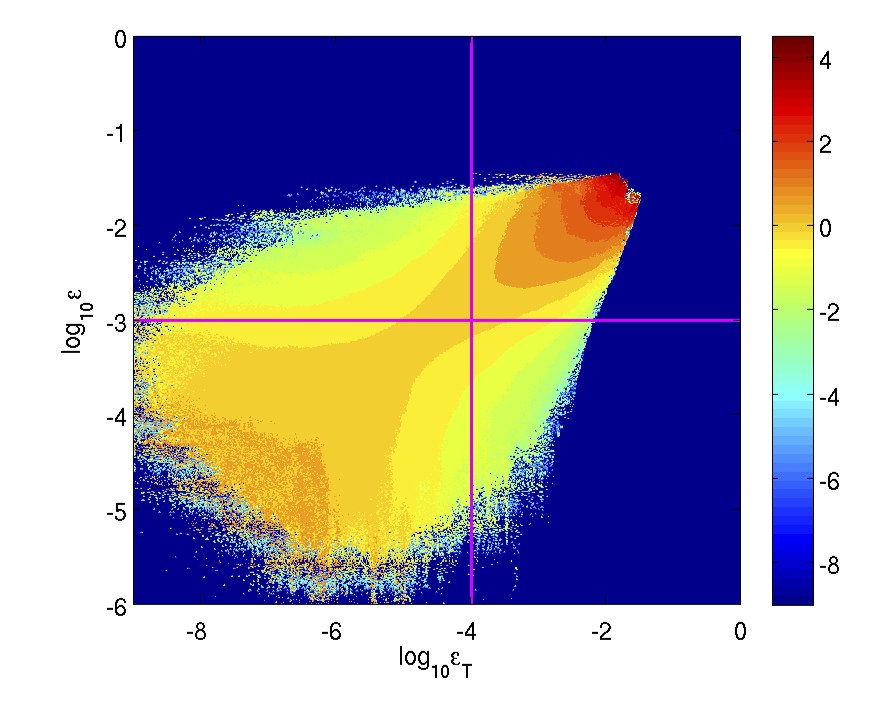}
\caption{Joint statistics of both dissipation rates $\Pi(\epsilon_T,\epsilon)$ as given by (\ref{joint1}) for SEM7 and SEM7a. Left: $Pr=0.7$. 
Right: $Pr=6$. The vertical and horizontal lines indicate the corresponding mean dissipation rates. Data are obtained in $V_b$. The color 
scale is given in decadic logarithm.}
\label{pdfjoint}
\end{figure}
%---------------------------------------------------------------------

\subsection{Joint Statistics}

The joint statistics of both dissipation rates is displayed in Fig. \ref{pdfjoint}. We show the joint and normalized probability density function which is given by 
%---------------------------------------------------------------------
\begin{equation}
\Pi(\epsilon_T,\epsilon)=\frac{P(\epsilon_T,\epsilon)}{P(\epsilon_T) P(\epsilon)}\,.
\label{joint1}
\end{equation}
%---------------------------------------------------------------------
The contour levels are plotted in logarithmic values as indicated by the color bar.  Similar to \cite{Kaczorowski2013} for RB convection or to \cite{Hamlington2012a} for a channel flow,  the support of $\Pi$ has an ellipsoidal form  with a tip at the joint high-amplitude events.  The joint PDF $P(\epsilon_T,\epsilon)$ is here normalized by the corresponding
single quantity PDFs, $P(\epsilon)$ and $P(\epsilon_T)$. In this way we highlight the correlations between both fields. If $\Pi(\epsilon_T,\epsilon)$ is larger than unity then the correlation is larger than if the two 
dissipation rate fields were statistically independent. It can be observed that 
 the support of the joint PDF for $Pr=6$ is shifted to smaller amplitudes in comparison to $Pr=0.7$, which is in agreement with Fig. \ref{pdfPrandtl}. 
In both cases, the high-amplitude events are correlated strongest, exceeding the corresponding mean amplitudes by at least two orders of magnitude. 
%---------------------------------------------------------------------

\section{Summary and discussion}
We have computed both global and local measures of dissipation and heat transport from high resolution direct 
numerical simulations of turbulent Rayleigh-B\'enard convection using a spectral element method. We find that 
the global measures of heat transport, such as Nusselt number, time-averaged temperature profiles, and volume-averaged 
dissipation rates, are fairly insensitive to insufficient resolution, as long as the mean Kolmogorov length is resolved. 
However,  if one computes instead plane-averaged or even more local dissipation rates, one finds that the Gr\"otzbach 
criteria (or something even more stringent as in (\ref{etaKz22})) needs to be satisfied for every grid point in order to have the system properly resolved. The main effects of a 
poorly resolved simulation are that some of the largest dissipation (both thermal and viscous) scales in the system are 
not resolved, especially in the bulk where the computational grid is coarsest. Our investigations suggest that
the refined SEM analysis which we conducted to study the statistics of dissipation fields require at least
%---------------------------------------------------------------------
\begin{equation}
\frac{\Delta z(z)}{\langle\eta_K(z)\rangle_{A,t}}\lesssim \frac{\pi}{2} \;\;\;\;\mbox{for}\;\;\;\; Pr\gtrsim 1\,.
\label{etaKz22}
\end{equation}
%---------------------------------------------------------------------
This follows e.g. from the data displayed in Figure \ref{pdf3} for the largest Rayleigh number. It is 
clear that such a criterion can be applied a posteriori only. Recall also that the horizontal spacing was always
finer in the present cases such that a geometric mean remains smaller than $\Delta z(z)$.

We have also compared our SEM results with a FDM code and find excellent agreement for global quantities 
such as Nusselt number and temperature profiles, and even fair agreement with globally averaged dissipation rates. 
The only discrepancy is with the vertical profiles of the mean  dissipation rates, which  disagree by about 9\%.

Once we determined our resolution criteria, we then compared local dissipation rates ($\epsilon, \epsilon_T$) and 
the local dissipation scale $\eta_k$ as a function of Rayleigh number for our sufficiently-resolved simulations. Local dissipation
scales can be considered as a generalization of the mean Kolmogorov dissipation scale which incorporate the spatially intermittent 
nature of the energy dissipation field. Local scales below the Kolmogorov scales are related to strong local gradients or high-amplitude 
dissipation events. We find that the local dissipation scales in the entire cell have a wider range of values than the dissipation scales in the bulk 
of the cell. But in all cases, there is a fairly wide range of dissipation scales both above and below the mean Kolmorgov dissipation scale. 
The range of these local scales is a manifestation of the  intermediate dissipation range (IDR) which exists in the crossover region between the inertial and viscous range. The IDR was developed
in the multifractal formalism \cite{Paladin1987,Nelkin1990,Vergassola1991,Biferale2008}. Similar to previous studies in box turbulence and 
channel flow turbulence, this range increases as the Rayleigh number grows. We have  shown here that the dissipation scales on the left end of the PDFs become smaller as Rayleigh number 
increases, and correspondingly the probability of largest dissipation scales decreases.
We also found, by looking at the fourth 
moment of the thermal dissipation rate, that high-amplitude but rare dissipation events can dominate the tails of the 
PDFs of the thermal dissipation rates. This highlights the sensitivity of turbulent RBC to such rare, but extreme events 
and calls for caution when generalizing statistical quantities in turbulent RBC.
 We also computed the joint statistics of the kinetic energy and thermal dissipation rates and find that the high amplitude events are the most strongly correlated.

Finally we compared results at two different Prandtl numbers. 
The range of local dissipation scales 
becomes smaller when $Pr>1$ which is in line with smaller amplitudes of both dissipation rates.  Our estimates (\ref{diss1b}) indicate that the resolution demands grow significantly when the Prandtl number 
is decreased starting from $Pr\approx 1$.  Equation (\ref{diss1b}) suggests a stronger Prandtl number dependence on the dissipation scales, namely 
$\langle\eta_K\rangle\sim Pr^{3/8}$, for cases decreasing from $Pr\approx 1$ to $Pr\ll 1$ than for those which increase from $Pr\approx 1$ 
to $Pr\gg 1$ (where $\langle\eta_B\rangle\sim Pr^{-1/8}$).  On the numerical side, a second challenge appears that is related to the high diffusivity of temperature field and which has 
been discussed recently for the case of passive scalar mixing at very low Schmidt number \cite{Yeung2013}. An explicit time advancement becomes 
increasingly demanding since the scalar relaxes increasingly faster. Preliminary studies suggest  e.g. that for the Prandtl number of mercury 
$(Pr=0.021)$ a mesh is necessary that equals the one which we used for $Pr=0.7$ for a Rayleigh number larger by a factor of one hundred. 

\ack First, we would like to thank Paul F. Fischer for his continuous help in getting the nek5000 spectral element software package optimized and adapted 
to our Rayleigh-B\'{e}nard convection problem. J. Schumacher thanks the Deutsche Forschungsgemeinschaft for financial support within the 
Research Unit FOR 1182 and the German-Israeli-Foundation with Grant 1072-6.14/2009. Supercomputing resources on Blue Gene/Q Juqueen 
at the J\"ulich Supercomputing Centre have been obtained within Grant HIL07 which has been selected as a large-scale project in the German 
Gauss Centre for Supercomputing. We thank them for this steady support of our work. J. D. Scheel acknowledges also an INCITE director's discretionary allocation for 
 Blue Gene/P Intrepid and Blue Gene/Q Mira at Argonne National Laboratory.  The work benefited from discussions with Bruno Eckhardt, 
Roberto Verzicco and Katepalli R. Sreenivasan.

\clearpage

\section*{Appendix. Additional resolution studies}
\subsection*{Sensitivity with respect to vertical element spacing}
\begin{figure}
\centering
\includegraphics[width=0.65\textwidth]{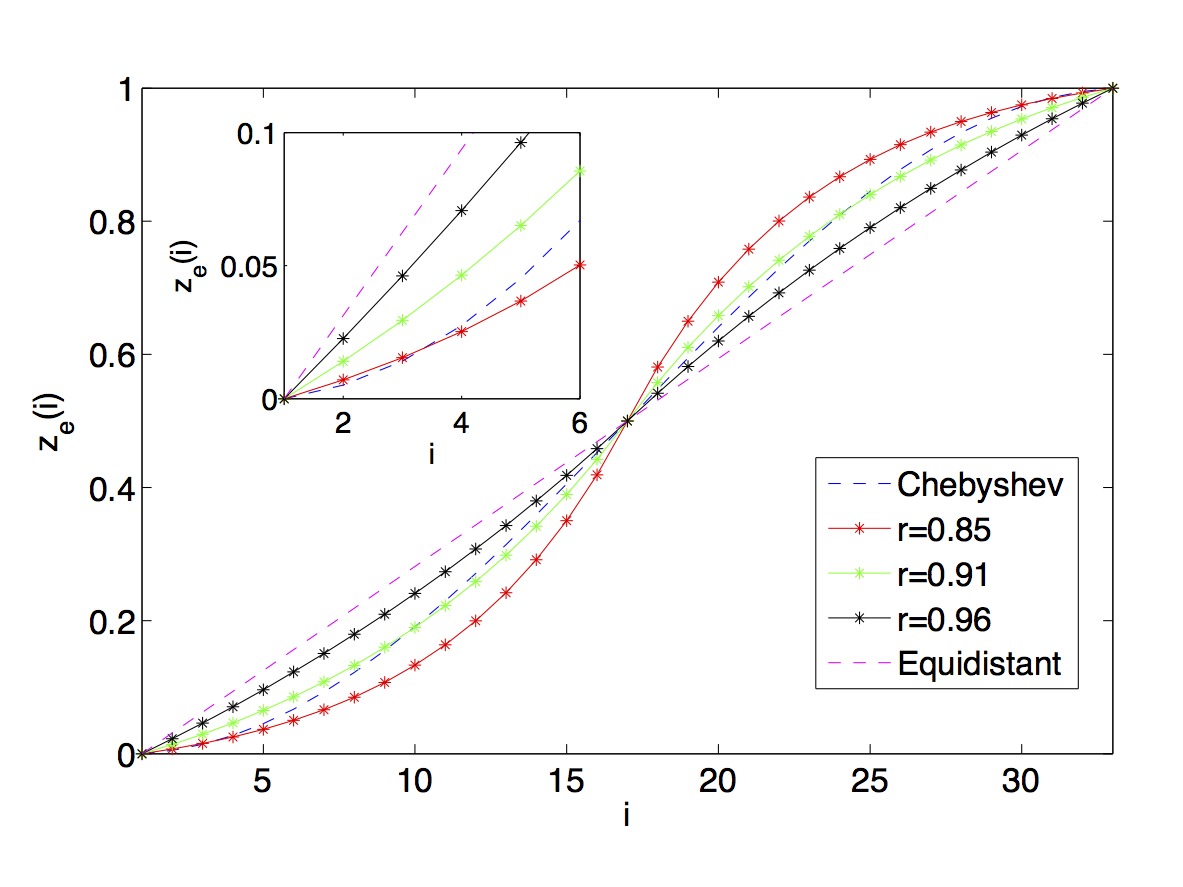}
\caption{Display of the vertical primary node structure with $N_{e,z}=32$, where $i=1\dots N_{e,z}+1$ and $z_e(i)\in [0,1]$. The geometric scaling 
factor $r$ causes different narrow node spacings in the vicinity of the top and bottom plate. For visual comparison we add an equidistant spacing and 
the Chebyshev collocation points. The inset displays a zoom into the vicinity of the bottom plate. }
\label{Cheby}
\end{figure}

In this section, we describe in brief one way to obtain an optimal scaling factor $r$ given by (\ref{geometric}). 
As $r$ becomes smaller, the elements become more clustered towards the boundary plates as shown in Figure 
\ref{Cheby}. Table \ref{Tab3} summarizes ten different test runs at fixed $Ra$, $Pr$ and $\Gamma$. In all cases the horizontal mesh 
(see again Figure \ref{grid_0}) and the total number of elements $N_e$ remain unchanged. We varied the polynomial order $N$ and $r$
only.
 
\begin{figure}
\centering
\includegraphics[width=0.7\textwidth]{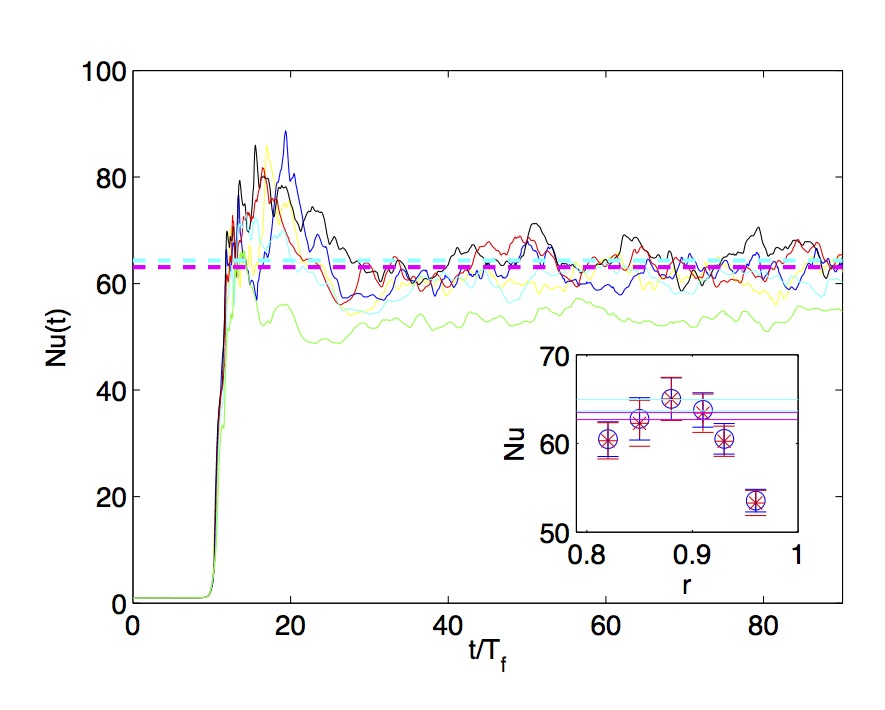}
\includegraphics[width=0.7\textwidth]{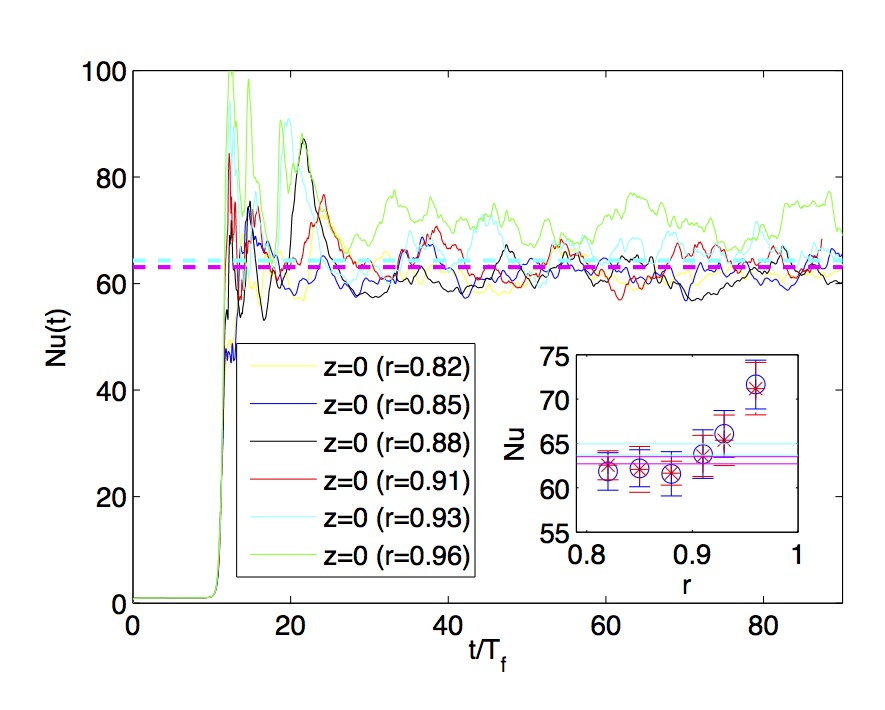}
\caption{Nusselt number at $z=0$ versus time. Results at differently stretched vertical grids are shown. The dashed lines show the 
reference values of $Nu$ from other DNS:  cyan dashed line for a run at same $Ra$ from \cite{Bailon2010} and the magenta dashed 
line from \cite{Wagner2012}. The number  of elements is the same in both series. Top: polynomial order $N=3$ with runs T1, T3, T5, 
T7 and T9. Bottom: $N=5$ with runs T2, T4, T6, T8 and T10. The insets in both figures display $Nu$ as obtained by a time average 
at $z=0$ (blue circles) and 1(red stars) as well as the corresponding error bars. For comparison we add again $Nu\pm \sigma$ from 
\cite{Bailon2010,Wagner2012} in the same color style as in the main figures. All data are for $Ra=10^9$ and $\Gamma=1$  (see 
Table \ref{Tab3}).}
\label{res2}
\end{figure}

Figure \ref{res2} shows time series of the Nusselt numbers for the different values of $r$ obtained at $z=0$, i.e., $Nu(t)=-\partial\langle 
T\rangle_{A}/\partial z|_{z=0}$. The values $Nu(t)$ fluctuate about their temporal means. These fluctuations do not decrease when $N$ 
is increased, i.e. when the resolution is improved (not shown).  A systematic effect for an increase of $r$ is clearly visible in the insets of 
both figures, where we report the time averages of $Nu(t)$ at both plates with the error bars corresponding to the standard deviation $\sigma$.  Neither an equidistant nor a 
strongly non-uniform grid are preferable since they give the largest discrepancy in Nusselt number. There is a trade-off between resolving the boundary layers (non-uniform grid) and the bulk (equidistant grid). Based on our analysis here, scaling factors of about $r\approx 0.9$ seem to be the optimum and were kept for the 
rest of the studies. For the present studies, we have chosen $r=0.91$ and have also matched finer primary element meshes correspondingly.
%---------------------------------------------------------------------
%-----------------------------------------------------------------
\begin{table}
\begin{center}
\begin{tabular}{cccccc}
\hline
Run & $N$ & $r$ &  $N_{e}$ & $N_{e,z}$ & $N_z$\\
\hline
T1 & 3 & 0.85 & 30720 & 32 &96\\
T2 & 5 & 0.85 & 30720 & 32 &160\\
T3 & 3 & 0.88 & 30720 & 32 &96\\
T4 & 5 & 0.88 & 30720 & 32 &160\\
T5 & 3 & 0.91 & 30720 & 32 &96\\
T6 & 5 & 0.91 & 30720 & 32 &160\\
T7 & 3 & 0.93 & 30720 & 32 &96\\
T8 & 5 & 0.93 & 30720 & 32 &160\\
T9 & 3 & 0.96 & 30720 & 32 &96\\
T10 & 5 & 0.96 & 30720 & 32 &160\\
\hline
\end{tabular}  
\end{center}
\caption{Parameters of the different spectral element test runs T1 to T10 with different vertical spacing. We display the order $N$ of the 
Legendre polynomials,  the geometric stretching factor $r$ in the vertical direction,
the total number of spectral elements, $N_{e}$, the number of spectral elements with respect to $z$ direction, $N_{e,z}$, 
and the number of grid cells resulting from primary and secondary nodes with respect to $z$ direction, $N_z=N_{e,z} N$. In all cases, $Ra=10^9$, $\Gamma = 1$ and $\Pr = 0.7$.}
\label{Tab3}
\end{table}

%---------------------------------------------------------------------
%---------------------------------------------------------------------

\subsection*{Complementary series of resolution tests}
%---------------------------------------------------------------------
\begin{table}
\begin{center}
\begin{tabular}{ccccccccc}
\hline
Run & $N$ & $(N_{e}, N_{e,z})$ & $N_z$ & $Nu(0)\pm \sigma$ & $Nu(1)\pm \sigma$ & $Nu_V\pm \sigma$ & $\Lambda_T$ & $\Lambda_v$\\
\hline
T11 & 7 &  (30720, 32) &96 & 61.8$\pm 1.0$ & 61.8$\pm 1.1$ & 63.1$\pm 2.4$ & 1.7\% & 4.0\% \\
T12 & 7 & (256000, 64) & 448 & 62.8$\pm 1.0$ & 62.6$\pm 1.1$ & 62.9$\pm 4.7$ & 0.4\% & 0.2\%\\
T13 & 7 & (875520, 96) & 672 & 62.8$\pm 2.0$ & 62.9$\pm 2.3$ & 62.8$\pm 5.0$ & 0.1\% & 0.1\%\\
\hline
\end{tabular}  
\end{center}
\caption{Parameters of the different spectral element simulations T11 to T13. The three runs have different primary node meshes, 
but the same polynomial order $N=7$.  
We display the order $N$ of the Legendre polynomials,
the total number of spectral elements, $N_{e}$, the number of spectral elements with respect to $z$ direction, $N_{e,z}$, 
the number of grid cells resulting from primary and secondary nodes with respect to $z$ direction, $N_z=N_{e,z} N$, and the Nusselt numbers 
$Nu(z=0)$, $Nu(z=1)$ and $Nu_V$. Furthermore we list the relative errors $\Lambda_T$ and $\Lambda_v$ (see Eqns. (\ref{rel1a})). All runs 
are at $Ra=10^9$, $\Gamma=1$ and $Pr=0.7$. Note that T13 equals SEM9.}
\label{Tab5}
\end{table}

\begin{figure}
\centering
\includegraphics[width=1.0\textwidth]{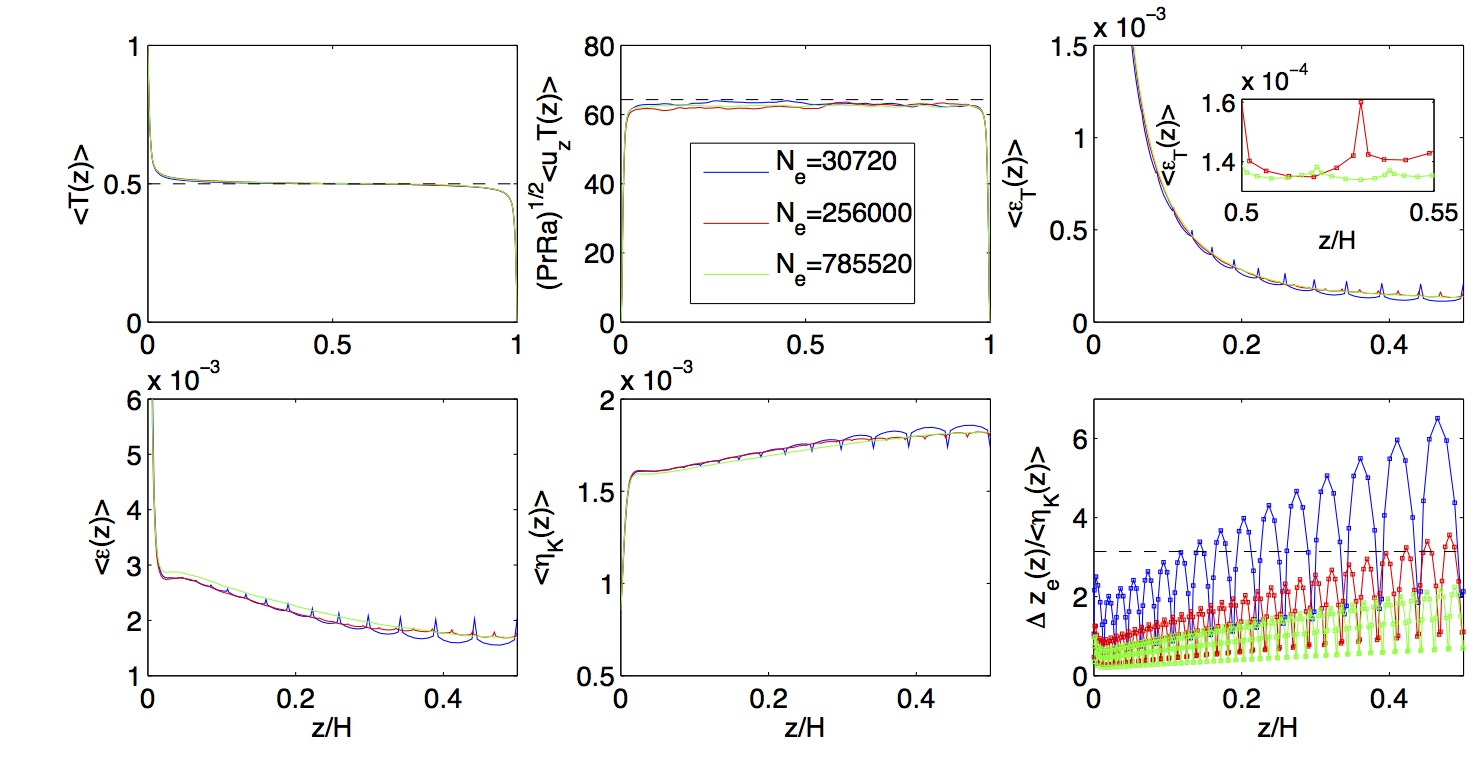}
\caption{Resolution tests for $Ra=10^9$, $\Gamma = 1$, and $N=7$ using different primary meshes  (runs T11 to T13 in  Table \ref{Tab3}).
We compare the same quantities as in Figure \ref{grid3a}. The dashed line in the lower right panel marks $\Delta z_e(z)/
\eta_K(z)=\pi$. The dashed line in the upper mid panel is the Nusselt number from run FDM1 as displayed in Table \ref{Tab2}. }
\label{grid3}
\end{figure}
%---------------------------------------------------------------------
A complementary series of resolution tests in comparison to those reported in Sec. 3.1 is presented in  Table \ref{Tab5} and Figure \ref{grid3}.
In the test runs T11 to T13 
we varied the primary element meshes and left the polynomial order of 
each element unchanged. The outcome from this series is similar to what was already demonstrated in the main text. While
the vertical profiles for the mean temperature or the mean convective  heat flux are practically equal, differences manifest 
for the gradient fields (see Figure \ref{grid3} for details).

\subsection*{Spatial derivatives in the spectral element method} 

In order to illustrate how we take very accurate derivatives in the SEM, we use as an example a one-dimensional case 
with the reference element $\Omega=[-1, 1]$. A spectral approximation of a function $u_e\in L_w^2(\Omega)$ (with $w(x)$ being a positive weight function) 
can be written as follows 
%----------------------------------------------------------------------
\begin{equation}
u_e(x)=\sum_{k=0}^N u(\xi_k) \psi_k(x)\,,
\label{expans1}
\end{equation}
%----------------------------------------------------------------------
where $\psi_k(x)$ is the $k$th order basis function and the (N+1) points are the nodes of the Gauss-Lobatto quadrature. They 
are determined by the Gauss-Lobatto integration theorem \cite{Deville2002}. For the approximation, one has to take a set of 
polynomials which form an orthogonal system of the underlying Hilbert space of square-integrable functions $L_w^2(\Omega)$.

The first derivative of the function $u_e(x)$  at the GLL points is
%----------------------------------------------------------------------
\begin{equation}
D_e u(\xi_k)=\sum_{j=0}^N u(\xi_j) \frac{d\psi_j(x)}{dx}\Big|_{x=\xi_k}\,.
\label{deriv1}
\end{equation}
%----------------------------------------------------------------------

\begin{figure}
\centering
\includegraphics[width=0.9\textwidth]{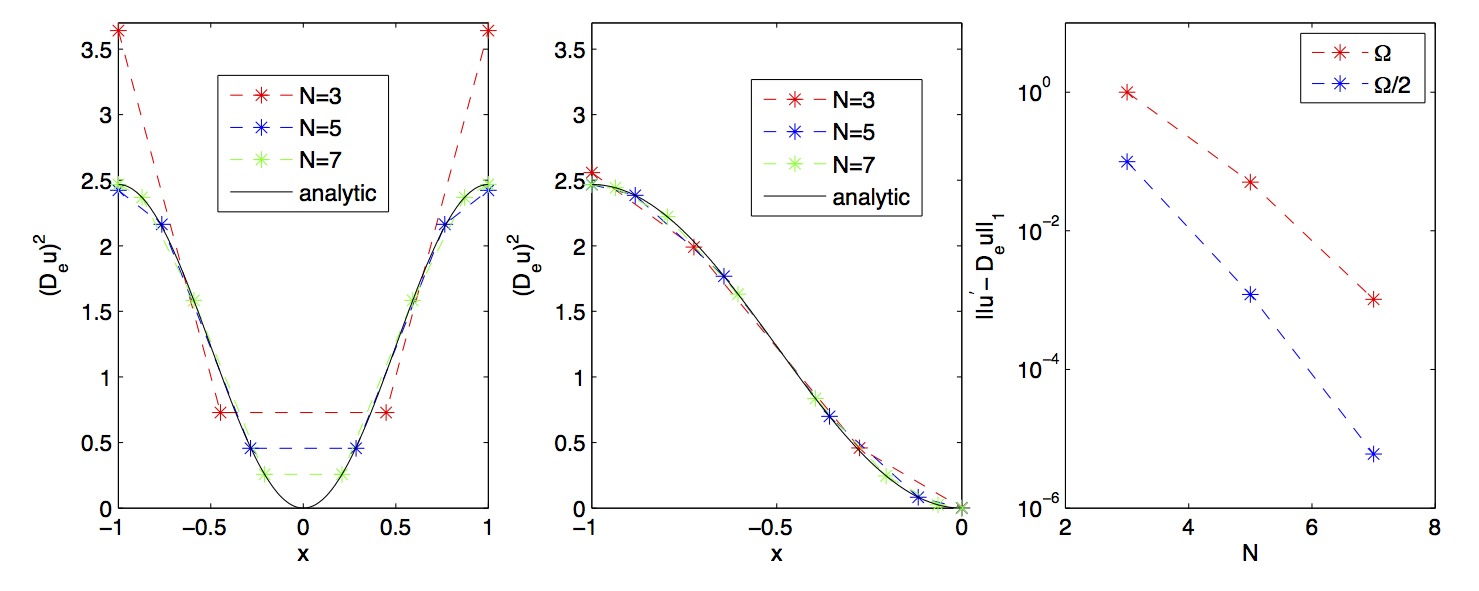}
\caption{First-order derivative as a function of polynomial order and element size. Left: the results for $(du/dx)^2$ where 
$u(x)=\cos(\pi x/2)$ with $x\in[-1,1]$ are displayed. Mid: same results but for an primary element mesh ($\Omega/2$) twice as fine as in the
left figure ($\Omega$). Right: exponential convergence of the error as defined in (\ref{deriv2d}).}
\label{der}
\end{figure}

Starting from Eq. (\ref{basefunctions}) together with the relation $(1-x^2)L^{\prime}(x)=0$ for $x=\xi_k$ and substituting the Legendre 
differential equation $((1-x^2) L^{\prime}(x))^{\prime}=-N(N+1)L_N(x)$ the derivative becomes
%----------------------------------------------------------------------
\begin{equation}
\frac{d\psi_j(x)}{dx}\Big|_{x=\xi_k}= \left\{ 
\begin{array}{l l} 
 \frac{L_N(\xi_k)}{(\xi_k-\xi_j)L_N(\xi_j)}  & \textrm{ if } \;\;\xi_k\ne \xi_j \\ \\
 \frac{N(N+1)}{4}                                          & \textrm{ if }  \;\;k=j=N \\ \\
-\frac{N(N+1)}{4}                                          & \textrm{ if }  \;\;k=j=0 \\ \\
0                                                                        &  \textrm{ otherwise}  
\end{array}  
\right.  
\label{loglaw}
\end{equation}
%----------------------------------------------------------------------
We also recall that $L_N(-1)=(-1)^N$ and $L_N(1)=1$ for all $N$. Thus, the derivative at the boundary $x=\xi_0$ is given by
%----------------------------------------------------------------------
\begin{equation}
D_e u(\xi_0)=-\frac{N(N+1)}{4} u(\xi_0) -\sum_{j=1}^{N} \frac{u(\xi_j)}{(\xi_0-\xi_j)L_N(\xi_j)}\,,
\label{deriv2a}
\end{equation}
%----------------------------------------------------------------------
at $x=\xi_k$ for $0<k<N$ by
%----------------------------------------------------------------------
\begin{equation}
D_e u(\xi_k)=\sum_{j=0\atop j\ne k}^{N} \frac{u(\xi_j)L_N(\xi_k)}{(\xi_k-\xi_j)L_N(\xi_j)}\,,
\label{deriv2b}
\end{equation}
%----------------------------------------------------------------------
and at $x=\xi_N$ by
%----------------------------------------------------------------------
\begin{equation}
D_e u(\xi_N)= \sum_{j=0}^{N-1} \frac{u(\xi_j)}{(\xi_0-\xi_j)L_N(\xi_j)}+\frac{N(N+1)}{4}u(\xi_N)\,.
\label{deriv2c}
\end{equation}
%----------------------------------------------------------------------
In Figure \ref{der} we summarize the results for a simple function $u(x)=\cos(\pi x/2)$. With a view to dissipation rates 
we are interested in the accuracy for quantities that contain $(du/dx)^2$. In the left panel of Figure \ref{der}, we compare the 
derivative as obtained from (\ref{deriv2a})--(\ref{deriv2c}). We see that the errors at the boundary result in strong overshoots
at the element boundary which are amplified by the second power of the derivatives as in the dissipation rates. The mid panel of Figure 
\ref{der} repeats the analysis for a primary element node mesh of half the size obtained here by $\tilde x\rightarrow (x-1)/2$. An 
increase in resolution of the primary element node mesh reduces the errors significantly. The error can be quantified by 
%----------------------------------------------------------------------
\begin{equation}
\parallel u^{\prime}-D_e u\parallel= \sum_{j=0}^{N} |u^{\prime}(\xi_j)-D_e u(\xi_j) |\,,
\label{deriv2d}
\end{equation}
%----------------------------------------------------------------------
which is shown in the right panel of Figure \ref{der}. The exponential convergence with respect to the polynomial order $N$ is 
clearly demonstrated in the right panel of the figure. Thus a combination of both increasing the polynomial order and the node 
mesh leads to an accurate calculation of spatial moments.     
%---------------------------------------------------------------------
%---------------------------------------------------------------------


\section*{References}
\begin{thebibliography}{100}
\bibitem{Ahlers2009}
Ahlers G, Grossmann S and Lohse D 2009
Heat transfer and large scale dynamics in turbulent Rayleigh-B\'{e}nard convection
{\it Rev. Mod. Phys.} {\bf 81} 503--537

\bibitem{Bailon2010}
Bailon-Cuba J,  Emran M S and Schumacher J 2010 
Aspect ratio dependence of heat transfer and large-scale flow  in turbulent convection
{\it J. Fluid Mech.} {\bf 655} 152--173

\bibitem{Batchelor1959}
Batchelor G K 1959
Small scale variation of convected quantities like temperature in a turbulent fluid
{\it J. Fluid Mech.} {\bf 5} 113--133

\bibitem{Biferale2008}
Biferale L 2008
A note on the fluctuations of dissipation scale in turbulence
{\it Phys. Fluids} {\bf 20} 031703 (4 pages)

\bibitem{Chilla2012} 
Chill\`{a} F and Schumacher J 2012 
New perspectives in turbulent Rayleigh-B\'{e}nard convection 
{\it Eur. J. Phys. E} {\bf 35} 58 (25 pages)

\bibitem{Deville2002}
Deville M O, Fischer P F and Mund E H 2002
{\em High-order methods for incompressible fluid flow}
(Cambridge, UK:  Cambridge University Press)

\bibitem{Emran2008}
Emran M S and Schumacher J 2008
Fine-scale statistics of temperature and its derivatives in convective turbulence 
{\it J. Fluid Mech.} {\bf 611} 13--34

\bibitem{Emran2012}
Emran M S and Schumacher J 2012
Conditional statistics of thermal dissipation rate in turbulent Rayleigh-B\'{e}nard convection
{\it Eur. J. Phys. E} {\bf 35}  108 (8 pages)

\bibitem{nek5000}
Fischer P F,  Lottes J W and Kerkemeier S G 2012 http://nek5000.mcs.anl.gov

\bibitem{Fischer1997}
Fischer P F 1997
An overlapping Schwarz Method for spectral element solution of the incompressible Navier-Stokes equations
{\it J. Comp. Phys.} {\bf 133} 84--101

\bibitem{Fischer2001}
Fischer P F  and Mullen J 2001
Filter-Based Stabilization of Spectral Element Methods
{\it Comptes Rendus de l'Acad\'emie des Sciences Paris, S\'erie I - Analyse num\'erique} {\bf 332} 265--270

\bibitem{Fischer2008}
Fischer P F, Lottes J W, Pointer D and Siegel A 2008
Petascale algorithms for reactor hydrodynamics
{\it J. Phys. Conf. Ser.} {\bf 125} 012076 (5 pages)

\bibitem{Vergassola1991}
Frisch U and Vergassola M 1991
A prediction of the multifractal model -- the intermediate dissipation range
{\it Europhys. Lett.} {\bf 14} 439--444

\bibitem{Hamlington2012}
Hamlington P E, Krasnov D, Boeck T and Schumacher J 2012
Local dissipation scales and energy dissipation-rate moments in channel flow  
{\it J. Fluid Mech.} {\bf 701} 419--429

\bibitem{Hamlington2012a}
Hamlington P E, Krasnov D, Boeck T and Schumacher J 2012
Statistics of the energy dissipation rate and local enstrophy in turbulent channel flow
{\it Physica D} {\bf 241} 169--177

\bibitem{Groetzbach1983}
Gr\"otzbach G 1983
Spatial resolution requirements for direct numerical simulation of the Rayleigh-B\'{e}nard convection
{\it J. Comp. Phys.} {\bf 49} 241--264

\bibitem{Kaczorowski2013}
Kaczorowski M and Xia K-Q 2013
Turbulent flow in the bulk of Rayleigh-B\'{e}nard convection: small-scale properties in a cubic cell
{\it J. Fluid Mech.} {\bf 722}  596--617

\bibitem{Mellado2010}
Mellado J P 2010
The evaporatively driven cloud-top mixing layer
{\it J. Fluid Mech.} {\bf 660} 5--36

\bibitem{Nelkin1990}
Nelkin M 1990
Multifractal scaling of velocity derivatives in turbulence
{\it Phys. Rev. A} {\bf 42} 7226--7229

\bibitem{Paladin1987}
Paladin G and Vulpiani A 1987
Degrees of freedom of turbulence
{\it Phys. Rev. A} {\bf 35} 1971--1973 

\bibitem{Pope2000}
Pope S B 2000
{\em Turbulent Flows}
(Cambridge, UK:  Cambridge University Press)

\bibitem{Scheel2012} 
Scheel J D, Kim E and White K R 2012 
Thermal and viscous boundary layers in  turbulent Rayleigh-B\'{e}nard convection 
{\it J. Fluid Mech.} {\bf 711} 281--305

\bibitem{Schumacher2005}
Schumacher J, Yeung P K and Sreenivasan K R 2005
Very fine structures in scalar mixing 
{\it J. Fluid Mech.} {\bf 531} 113--122

\bibitem{Schumacher2007}
Schumacher J 2007
Sub-Kolmogorov scale fluctuations in fluid turbulence 
{\it Europhys. Lett.}  {\bf 80}  54001 (6 pages)

\bibitem{Schumacher2007a}
Schumacher J, Sreenivasan K R and Yakhot V 2007
Asymptotic exponents from low-Reynolds-number flows
{\it New J. Phys.}  {\bf 9}  89 (19 pages)

\bibitem{Schumacher2010}
Schumacher J, Eckhardt B and Doering C R 2010
Extreme vorticity growth in Navier-Stokes turbulence
{\it Phys. Lett. A} {\bf 374} 861--865.

\bibitem{Shi2012}
Shi N, Emran M S and Schumacher J 2012
Boundary layer structure in turbulent Rayleigh-B\'{e}nard convection 
{\it J. Fluid Mech.} {\bf 706} 5--33

\bibitem{Shishkina2010}
Shishkina O, Stevens R A J M, Grossmann S and Lohse D 2010
Boundary layer structure in turbulent thermal convection and its consequences for the required numerical resolution
{\it New J. Phys.} {\bf 12} 075022 (17 pages)

\bibitem{Shraiman1990}
Shraiman B I and Siggia E D 1990
Heat transport in high-Rayleigh-number convection
{\it Phys. Rev. A} {\bf 42} 3650--3653.

\bibitem{Sreenivasan1997}
Sreenivasan K R and Antonia RA 1997 
The phenomenology of small-scale turbulence 
{\it Annu. Rev. Fluid Mech.} {\bf 29} 435--472

\bibitem{Sreenivasan2004}
Sreenivasan K R 2004 
Possible effects of small-scale intermitency in turbulent reacting flows 
{\it Flow Turb. Combust.} {\bf 72} 115--132

\bibitem{Stevens2010}
Stevens R A J M, Verzicco R and Lohse D 2010
Radial boundary layer structure and Nusselt number in Rayleigh-B\'{e}nard convection
{\it J. Fluid Mech.} {\bf 643} 495--507
 
\bibitem{Tufo2001}
Tufo H M and Fischer P F 2001
Fast parallel direct solvers for coarse grid problems
{\it J. Parallel Distr. Com.} {\bf 61} 151--177

\bibitem{Verzicco1996}
Verzicco R and Orlandi P 1996
A finite-difference scheme for three-dimensional incompressible flows in cylindrical coordinates
{\it J. Comp. Phys.} {\bf 123} 402--414

\bibitem{Verzicco2003}
Verzicco R and Camussi R  2003
Numerical experiments on strongly turbulent thermal convection in a slender  cylindrical cell
{\it J. Fluid Mech.} {\bf 477} 19--49

\bibitem{Wagner2012}
Wagner S, Shishkina O and Wagner C 2012
Boundary layers and wind in cylindrical Rayleigh-B\'{e}nard cells
{\it  J. Fluid Mech.} {\bf 697} 336--366

\bibitem{Wallace2009}
Wallace J M 2009
Twenty years of experimental and direct numerical simulation access to the velocity gradient tensor: What have we learned about turbulence?
{\it Phys. Fluids} {\bf 21}  021301

\bibitem{Watanabe2004}
Watanabe T and Gotoh T 2004
Statistics of a passive scalar in homogeneous turbulence
{\it New J. Phys.} {\bf 6} 40 (36 pages)

\bibitem{Yakhot2005}
Yakhot V and Sreenivasan K R 2005 
Anomalous scaling of structure functions and dynamic constraints on turbulence simulations 
{\it J. Stat. Phys.} {\bf 121} (5) 823--841

\bibitem{Yakhot2006} 
Yakhot V 2006 
Probability densities in strong turbulence
{\it Physica D} {\bf 215} 166--174

\bibitem{Yeung2013}
Yeung P K and Sreenivasan K R 2013
Spectrum of passive scalars of high molecular diffusivity in turbulent mixing 
{\it J. Fluid Mech.} {\bf 716} R14


\bibitem{Zhou2010}
Zhou Q and Xia K-Q 2010
Universality of local dissipation scales in buoyancy-driven turbulence
{\it Phys. Rev. Lett.} {\bf 104} 124301

\end{thebibliography}
\end{document}